\documentclass[prb,aps,twocolumn,noeprint,10pt]{revtex4-2}
\usepackage{amsmath,amssymb,amsfonts}
\usepackage{newtxtext,newtxmath}
\usepackage{xcolor}
\usepackage{graphicx}
\usepackage[colorlinks=true,citecolor=blue,linkcolor=blue,urlcolor=blue]{hyperref}

\graphicspath{{./figs/}}
\newcommand{\includetikz}[1]{\,\vcenter{\hbox{\includegraphics{#1.pdf}}}}
\newcommand{\llangle}{\left\langle\!\left\langle}
\newcommand{\rrangle}{\right\rangle\!\right\rangle}

\begin{document}

\title{Locating the four-state Potts point of the frustrated \texorpdfstring{$J_1$--$J_2$}{J1-J2} Ising model with Klein-bottle partition functions}

\author{Satoshi Morita}
\email[]{smorita@keio.jp}
\affiliation{Graduate School of Science and Technology, Keio University,
  Yokohama, Kanagawa 223-8522, Japan}
\affiliation{Keio University Sustainable Quantum Artificial Intelligence Center (KSQAIC),
  Keio University, Minato-ku, Tokyo 108-8345, Japan}

% \date{\today}

\begin{abstract}
  We investigate the phase transition of the frustrated $J_1$--$J_2$ Ising model on the square lattice using the bond-weighted tensor renormalization group (BWTRG) method, focusing on the location of the 4-state Potts point $g^*$, at which the Ashkin-Teller critical line terminates.
  Using the $\mathbb{Z}_2$ symmetry of the tensor network, we compute the partition functions on the torus with the periodic and antiperiodic boundary conditions separately.
  Their ratio determines the transition temperatures with high accuracy and shows a plateau at the transition, whose value is described by the $\mathbb{Z}_2$ orbifold conformal field theory.
  We also introduce boundary tensors that connect the two ends of the system through a reflection, which allow us to compute the partition functions on the Klein bottle with a finite aspect ratio, twisted by the global spin flip and by the exchange of the horizontal and vertical stripe states.
  Since these two operations are conjugate in the $S_4$ symmetry at the Potts point but not in the $D_4$ symmetry on the Ashkin-Teller critical line, the two Klein-bottle ratios coincide only at the Potts point.
  Combining the estimates from the torus and the Klein bottle, we obtain $0.639\lesssim g^*\lesssim0.650$.
\end{abstract}

\maketitle

%%%%%%%%%%%%%%%%%%%%%%%%%%%%%%%%%%%%%%%%%%%%%%%%%%%%%%%%%%%%%%%%%
\section{Introduction}
\label{sec:intro}

Frustration, which arises when competing interactions cannot be satisfied simultaneously, is one of the central issues in statistical physics~\cite{diep2020frustrated}.
It not only suppresses long-range order but can also change the nature of the phase transition: the order of the transition and its universality class can depend on the strength of the competing interactions.
A typical example is the $J_1$--$J_2$ Ising model on the square lattice, in which the ferromagnetic nearest-neighbor interaction $J_1$ competes with the antiferromagnetic next-nearest-neighbor interaction $J_2$.
Despite its simplicity, the nature of its phase transition has been debated for a long time.

The ground state is ferromagnetic for $g\equiv|J_2|/J_1<1/2$ and is the fourfold-degenerate stripe state for $g>1/2$.
While the transition into the ferromagnetic phase belongs to the two-dimensional Ising universality class, the nature of the transition into the stripe phase has been controversial.
Monte Carlo (MC) studies have suggested the following picture~\cite{jin2012ashkinteller,kalz2012location,jin2013phase}:
for large $g$, the transition is continuous and belongs to the Ashkin-Teller (AT) universality class~\cite{ashkin1943statistics,delfino2004universal}, whose critical exponents vary continuously with $g$, whereas for $1/2<g<g^*$ it is weakly first order.
The AT critical line terminates at the 4-state Potts point at $g=g^*$.

However, the location of $g^*$ is still under debate.
For example, MC studies estimated $g^*\approx0.67$~\cite{jin2012ashkinteller,kalz2012location,jin2013phase}, $g^*\lesssim0.62$~\cite{li2024comprehensive}, and $g^*=0.696(10)$~\cite{sun2026relevantmarginal}.
The difficulty in determining $g^*$ originates from the nature of the transition near the Potts point.
For $g$ slightly below $g^*$, the first-order transition is so weak that the correlation length is extremely long, and it is hard to distinguish from a continuous transition in finite systems.
Moreover, the marginal operator at the Potts point causes logarithmic corrections, and the finite-size corrections decay very slowly.
Therefore, much larger system sizes than those accessible to conventional methods are required to locate $g^*$.

Tensor-network (TN) methods provide another approach to this problem, which can treat much larger systems than MC simulations.
In real-space renormalization methods, such as the tensor renormalization group (TRG)~\cite{levin2007tensor} and the higher-order TRG (HOTRG)~\cite{xie2012coarsegraining}, the linear system size grows exponentially with the number of renormalization steps.
In the transfer-matrix methods based on matrix product states (MPS), the system is treated directly in the thermodynamic limit.
In both approaches, the computational cost is polynomial in the bond dimension $\chi$, which controls the accuracy.
These methods have also been applied to this model.
Gangat and Kao located the phase boundary from the entanglement entropy along the TRG flow~\cite{gangat2019phase}.
Yoshiyama and Hukushima estimated $g^*\approx0.58$ from the discontinuity of the internal energy obtained by HOTRG~\cite{yoshiyama2023higherordera}.
Li and Yang concluded from the central charge, the scaling dimensions, and the Klein-bottle entropy that the transition is continuous down to $g\approx0.54$ and that the end point is a tricritical Ising point rather than the 4-state Potts point~\cite{li2021tensor}.
Gangat argued from the finite-entanglement scaling of MPS that the transition is weakly first order for any finite $g>1/2$ and becomes continuous only in the limit $g\to\infty$, i.e., $g^*=\infty$~\cite{gangat2024weak}.
Thus, the TN studies have reached no consensus on the location of $g^*$, nor even on the nature of the end point.

In this paper, we use the bond-weighted TRG (BWTRG)~\cite{adachi2022bondweighted,morita2025multiimpurity}, which significantly improves the accuracy of TRG by introducing weights on the bonds with only a small additional cost.
We employ a formulation of BWTRG that preserves the $90^\circ$ rotational symmetry of the lattice~\cite{lyu2026latticea}.
This is in contrast to HOTRG used in the previous studies, in which the coarse-graining is performed alternately in the horizontal and vertical directions and the rotational symmetry is not preserved.
The rotational symmetry is essential for the present model, since the horizontal and vertical stripe states are related to each other by the lattice rotation.
Moreover, since TN methods give the partition function itself, the partition functions with different boundary conditions can be computed directly.
At a critical point, their ratios converge to universal values, which are given by the modular-invariant partition functions of the conformal field theory (CFT)~\cite{ginsparg1988applied,difrancesco1997conformal,henkel1999conformal}.
Using the $\mathbb{Z}_2$ symmetry of the TN, we compute the partition functions on the torus with the periodic and antiperiodic boundary conditions separately, and their ratio gives both the transition point and the universality class.

On the AT critical line, however, the universal values of the ratios vary continuously with $g$, since the line is described by the $\mathbb{Z}_2$ orbifold CFT with a continuously varying compactification radius~\cite{saleur1987partition,ginsparg1988curiosities}.
Therefore, the ratios on the torus alone can locate the Potts point only through comparison with the universal values of the 4-state Potts CFT.
Here, in addition to the ratios on the torus, we focus on the emergent symmetry at the Potts point.
The four stripe states are related to each other by the dihedral group $D_4$, which is enhanced to the permutation group $S_4$ of the four states at the Potts point.
In $D_4$, the global spin flip and the exchange of the horizontal and vertical stripe states belong to different conjugacy classes, whereas they become conjugate in $S_4$.
While the spin flip acts on each spin and can be introduced as the antiperiodic boundary condition, the exchange is not an on-site operation on the spins but is realized by the reflection about a diagonal of the lattice.
Hence, the twist by the exchange cannot simply be introduced on the torus, but it can be introduced on the Klein bottle whose boundary condition involves the diagonal reflection.
We therefore compare two Klein bottles: one with the reflection about a lattice axis and the antiperiodic boundary condition, which forces a domain wall between a stripe state and its spin-flipped state, and the other with the diagonal reflection, which forces a domain wall between the horizontal and vertical stripe states.
These two kinds of domain walls are distinct in $D_4$ but become equivalent in $S_4$.
Thus, the two Klein-bottle ratios coincide only at the Potts point, which allows us to locate it without relying on the universal value of the 4-state Potts CFT.

Partition functions on the Klein bottle have also been used to characterize critical points.
Tu showed that the logarithm of the ratio of the partition functions on the Klein bottle and the torus gives a universal entropy determined by the CFT in the limit of a large aspect ratio~\cite{tu2017universal}.
This Klein-bottle entropy has been used to identify the CFTs of various critical points~\cite{tang2017universal,chen2017conformal,tang2019klein,zhang2023universal,wang2026kleina}, and also in the study of the present model mentioned above~\cite{li2021tensor}.
Recently, Shimizu and Ueda formulated the computation of the Klein-bottle partition function in TRG-based methods, in contrast to the previous MPS-based ones, by renormalizing the crosscap boundary states together with the bulk tensors with HOTRG~\cite{shimizu2026tensor}.
We instead introduce boundary tensors along the line on which the system is connected through the reflection (the reflection line).
These boundary tensors allow us to compute the Klein-bottle partition functions with a finite aspect ratio within BWTRG, including those twisted by internal symmetry operations, which are required for the comparison described above.

In this paper, we study the phase transition of the $J_1$--$J_2$ Ising model by computing the partition-function ratios on the torus and the Klein bottle with BWTRG for system sizes up to $L=2^{15}$.
The transition temperatures are determined from the ratio on the torus with much higher accuracy than the previous studies.
At the transition, the ratio on the torus shows a plateau over a wide range of $L$, and its crossing with the universal value of the 4-state Potts CFT gives an estimate $g=0.639$, which is likely to be a lower bound.
The crossing of the two Klein-bottle ratios, which does not rely on the universal values of the 4-state Potts CFT, gives an upper bound $g=0.650$.
The plateau value of the Klein-bottle ratio twisted by the exchange also crosses the universal value of the 4-state Potts CFT at $g=0.646$, which lies between these bounds.
From these estimates, we obtain $0.639\lesssim g^*\lesssim0.650$.

The rest of this paper is organized as follows.
In Sec.~\ref{sec:model-method}, we introduce the model, its TN representation, and the computation of the partition functions on the torus with BWTRG.
In particular, we explain in Sec.~\ref{sec:klein-bottle} why the Klein-bottle partition functions twisted by the two operations can detect the Potts point, based on the conjugacy classes of $D_4$ and $S_4$.
In Sec.~\ref{sec:results}, we present the phase diagram and the results of the partition-function ratios on the torus and the Klein bottle.
Section~\ref{sec:summary} is devoted to the summary and discussion.
The details of BWTRG, the partition functions of the $\mathbb{Z}_2$ orbifold CFT, and the transition temperatures are given in Appendices~\ref{appendix:bwtrg}, \ref{appendix:cft}, and \ref{appendix:tc}, respectively.

%%%%%%%%%%%%%%%%%%%%%%%%%%%%%%%%%%%%%%%%%%%%%%%%%%%%%%%%%%%%%%%%%
\section{Model and Method}
\label{sec:model-method}

%%%%%%%%%%%%%%%%%%%%%%%%%%%%%%%%%%%%%%%%%%%%%%%%%%%%%%%%%%%%%%%%%
\subsection{Frustrated Ising model}
\label{sec:model}

The frustrated $J_1$--$J_2$ Ising model on the square lattice is defined by the Hamiltonian,
\begin{equation}
  H = -J_1 \sum_{\langle i,j\rangle} S_i S_j
  - J_2 \sum_{\llangle i,j \rrangle} S_i S_j,
\end{equation}
where $S_i = \pm 1$ is the Ising spin at site $i$, and $\langle i,j\rangle$ and $\llangle i,j \rrangle$ denote the nearest-neighbor and next-nearest-neighbor pairs, respectively. We set $J_1=1$ as the energy unit and consider the case of the antiferromagnetic next-nearest-neighbor interaction, $J_2 < 0$, which introduces frustration into the system.

The ground state of the model is the ferromagnetic state for $g\equiv |J_2|/J_1 < 1/2$, and the stripe antiferromagnetic state for $g > 1/2$. At $g=1/2$, the system is highly frustrated, and the ground state is macroscopically degenerate.

The model exhibits a finite-temperature phase transition from the paramagnetic phase to the ordered phase except for the special case of $g=1/2$.
For $g<1/2$, the model exhibits a second-order phase transition belonging to the two-dimensional Ising universality class, as in the exactly solvable case of $g=0$.
For $g>1/2$, the nature of the phase transition is more complex and has been a subject of debate.
Many previous studies suggest that the phase transition is of first order for $1/2<g<g^*$, and of second order for $g\geq g^*$.

%%%%%%%%%%%%%%%%%%%%%%%%%%%%%%%%%%%%%%%%%%%%%%%%%%%%%%%%%%%%%%%%%
\subsection{TN representation and BWTRG}
\label{sec:tn-rep}

\begin{figure}[t]
  \centering
  \includegraphics{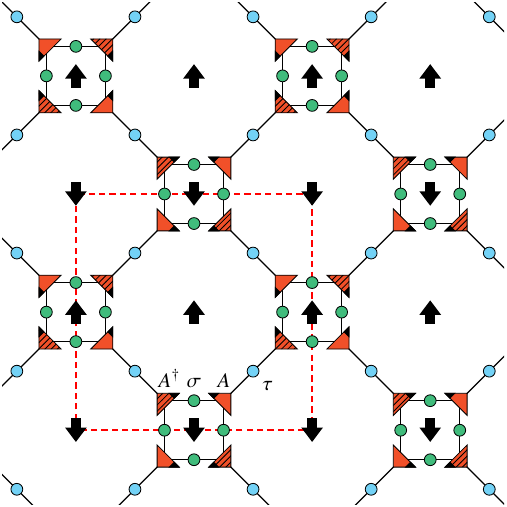}
  
  \caption{The initial TN representation of the frustrated Ising model on the square lattice. The black arrows indicate the location of spins and their directions correspond to the stripe state $H_+$ (see Sec.~\ref{sec:klein-bottle}).}
  \label{fig:stripe_lattice}
\end{figure}

The partition function of the model can be expressed as a TN in various ways.
In this study, we employ a triad TN with bond weights depicted in Fig.~\ref{fig:stripe_lattice}.
As described below, this representation is obtained by decomposing a bond-weighted TN defined on the dual square lattice~\cite{zhao2010renormalization}.
There are two kinds of bond weights, $\sigma$ and $\tau$, and the former is associated with the nearest-neighbor interaction $J_1$ and the latter with the next-nearest-neighbor interaction $J_2$.
We refer to $\sigma$ as the inner weight and $\tau$ as the outer weight as in Ref.~\cite{morita2025multiimpurity}.
A three-index tensor, $A$, represented by a triangle, is an isometry that connects the inner and outer weights.

The initial inner weight $\sigma$ is located between a pair of nearest-neighbor spins, and its basis corresponds to the product of the spin pair. That is, $\sigma$ is a $2\times2$ diagonal matrix with diagonal elements $e^{\pm \beta J_1}$, where $\beta=1/T$ is the inverse temperature.

The initial outer weight $\tau$ and the initial isometry $A$ are obtained from the eigenvalue decomposition of a four-leg tensor on the dual lattice.
Consider a square plaquette of the original lattice with spins at the vertices labeled as $S_1$ to $S_4$.
The four-leg tensor $T$ located at the dual lattice is defined by
\begin{equation}
  T_{ruld} = \exp\left[\beta J_2 (S_1 S_3 + S_2 S_4)\right],
\end{equation}
where $r,u,l,d$ correspond to the products of the spins on the right, up, left, and down edges, respectively.
Half of the elements of $T$ that do not satisfy the condition $(S_1S_2)(S_2S_3)(S_3S_4)(S_4S_1)=1$ are zero.
This tensor has the symmetry under 90-degree rotation of the lattice, $T_{ruld}=T_{uldr}$, and the symmetry under reflection, $T_{ruld}=T_{rdlu}$.
When reshaped into a matrix $T_{(rd),(ul)}$, it can be regarded as a symmetric matrix.
By the eigenvalue decomposition of this matrix, we can obtain the outer weight $\tau$ and the isometry $A$ as
\begin{equation}
  T_{ruld} = \sum_{s=1}^4 A_{rd,s} \tau_s A_{ul,s},
\end{equation}
or diagrammatically,
\begin{equation}
  \includetikz{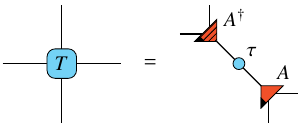}.
\end{equation}
The isometry $A$ is determined by the eigenvectors of $T_{(rd),(ul)}$ and satisfies the following identity:
\begin{equation}
  \includetikz{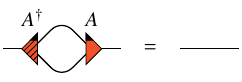}.
  \label{eq:iso_cond}
\end{equation}
We mark the triangle representing $A$ to indicate the order of its indices, since $A_{rd,s}\neq A_{dr,s}$.
The diagonal elements of $\tau$ are given by $e^{\beta J_2} \pm 1$ and $1 \pm e^{-\beta J_2}$, two of which are negative for $J_2 < 0$.
Thus, unlike the original BWTRG, the bond weights can take negative values.

Four equivalent decompositions of $T$ are possible due to its lattice-rotational symmetry.
We choose the decomposition that preserves the rotational symmetry around the center of the unit cell shown in Fig.~\ref{fig:stripe_lattice}, following Ref.~\cite{lyu2026latticea}.

Contracting the TN yields the partition function.
However, in exact calculations the dimension of the tensors grows exponentially, making approximate calculations necessary.
To address this, we employ the bond-weighted tensor renormalization group (BWTRG) method~\cite{adachi2022bondweighted}.
We extend the triad form of BWTRG~\cite{morita2025multiimpurity} to be applicable to TNs with lattice-rotational symmetry, as proposed in Ref.~\cite{lyu2026latticea}.

In each step of BWTRG, the outer weight $\tau$ becomes the inner weight $\tilde{\sigma}=\operatorname{sgn}(\tau)\left|\tau\right|^{k}$ in the next step, where the sign of $\tau$ is kept since $\tau$ can take negative values, and $k$ is a hyperparameter of BWTRG.
The outer weight $\tilde{\tau}$ and the isometry $\tilde{A}$ in the next step are obtained from the truncated eigenvalue decomposition of a small plaquette, in which we retain at most $\chi$ eigenvalues with the largest absolute values.
The details of the update are given in Appendix~\ref{appendix:bulk_tensor}.

After the above procedure, we obtain a new TN with the same structure as the original one, but with renormalized tensors.
A single step of BWTRG coarse-grains the network by a scaling factor of $\sqrt{2}$ with a 45-degree rotation.

The TN used here has a block structure as a $\mathbb{Z}_2$ symmetric TN~\cite{singh2010tensor,singh2011tensor}, corresponding to the constraint $(S_1S_2)(S_2S_3)(S_3S_4)(S_4S_1)=1$.
The $\mathbb{Z}_2$ charge here is not due to the spin inversion symmetry, but rather due to the conservation of the parity of domain walls.
This block structure is important for obtaining the partition function corresponding to the spin model, as described in the next subsection.

%%%%%%%%%%%%%%%%%%%%%%%%%%%%%%%%%%%%%%%%%%%%%%%%%%%%%%%%%%%%%%%%%
\subsection{Partition function on a torus}
\label{sec:partition-function}

The partition function on a torus can be obtained by taking the trace of the renormalized tensors.
Except for the bisection described in Sec.~\ref{sec:results}, we use a $\sqrt{2}\times\sqrt{2}$ cluster of the unit cell to calculate the partition function, which is represented by the following diagram:
\begin{equation}
  Z^\text{TN} = \includetikz{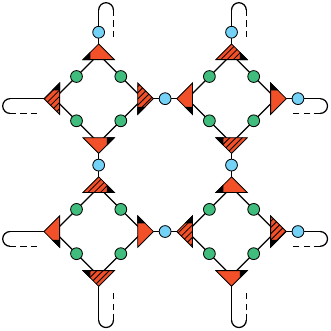},
  \label{eq:trace_16a}
\end{equation}
where the dashed lines represent the bonds connecting across the boundaries.
The system size corresponding to the TN after $n$ steps depends on the initial tensors.
For the TN in Sec.~\ref{sec:tn-rep}, the unit cell in Fig.~\ref{fig:stripe_lattice} corresponds to $2\times2$ spins at $n=0$ and to $L=2^{(n+2)/2}$ after $n$ steps of BWTRG.
Thus, Eq.~\eqref{eq:trace_16a} corresponds to $L=2^{(n+3)/2}$.
The computational cost of this contraction is proportional to $\chi^6$, the same as the trace of a single unit cell in Fig.~\ref{fig:stripe_lattice}, while the system size is larger by a factor of $\sqrt{2}$.
When locating the transition point by bisection, we compute the partition-function ratio from a smaller cluster with the $O(\chi^5)$ computational cost [Eq.~\eqref{eq:trace_4a} in Appendix~\ref{appendix:bulk_tensor}].

A contraction of the TN defined in Sec.~\ref{sec:tn-rep} does not correspond to the partition function of the original spin model on a torus.
This is because the tensors are defined on the dual lattice, and their indices represent the presence or absence of domain walls rather than the orientation of spins.
Under periodic boundary conditions, the number of domain walls wrapping around the torus is even.
However, our TN allows for configurations with an odd number of domain walls wrapping around the torus.
Such a configuration corresponds to a spin configuration with antiperiodic boundary conditions in the original spin model.
Therefore, a contraction of our TN on a torus, $Z^{\text{TN}}$, corresponds to the summation of the partition functions of the original spin model with periodic and antiperiodic boundary conditions,
\begin{equation}
  Z^{\text{TN}} = \frac{1}{2}(Z_\text{pp}+Z_\text{pa}+Z_\text{ap}+Z_\text{aa}).
  \label{eq:ZTN}
\end{equation}
Here, $Z_\text{pp}$ is the partition function of the original spin model with periodic boundary conditions in both directions, $Z_\text{pa}$ is the partition function with periodic boundary conditions in the horizontal direction and antiperiodic in the vertical direction, and so on.
The factor of $1/2$ arises because there are two spin configurations corresponding to each domain wall configuration.

Conversely, by fixing the total $\mathbb{Z}_2$ parity of the bonds crossing the boundaries, we can extract the value corresponding to the partition function of the original spin model.
For example, consider the case where the total $\mathbb{Z}_2$ parity of the bonds connecting in the horizontal direction is odd, and the bonds connecting in the vertical direction are even.
In this case, the number of domain walls wrapping around the horizontal direction is odd.
Since the spins are flipped when wrapping around the vertical direction, the contraction with this constraint corresponds to the partition function of the original spin model with antiperiodic boundary conditions in the vertical direction and periodic ones in the horizontal direction.
Therefore, we have the following correspondence:
\begin{equation}
  Z_\text{pp} = 2Z^{\text{TN}}_{++}, \quad 
  Z_\text{pa} = 2Z^{\text{TN}}_{-+}, \quad
  Z_\text{aa} = 2Z^{\text{TN}}_{--},
\end{equation}
where the subscripts of $Z^{\text{TN}}$ indicate the total $\mathbb{Z}_2$ parities of the bonds connecting in the horizontal and vertical directions.
In our construction, the lattice-rotational symmetry of the TN implies $Z_\text{ap} = Z_\text{pa}$.

The ratio of the partition functions with different boundary conditions is an important quantity to detect phase transitions.
In this paper, we consider the following ratio of partition functions:
\begin{equation}
  R_\text{pa} \equiv \frac{Z_\text{pa}}{Z_\text{pp}}.
\end{equation}
The logarithm of this ratio corresponds to the free energy of domain walls as $-\beta F_{\text{dw}}=\ln R_\text{pa}$.
In the ordered phase, creating a domain wall costs a free energy proportional to the system size $L$, and thus $R_\text{pa}$ converges to zero in the thermodynamic limit.
On the other hand, in the disordered phase, the free energy difference due to boundary conditions becomes exponentially small with respect to $L$, and thus $R_\text{pa}$ converges to one.
Therefore, the phase at a given temperature can be identified by whether $R_\text{pa}$ converges to zero or one as the system size $L$ increases, and the transition point $T_\text{c}$ is located as the boundary between the two behaviors.

Exactly at criticality, $R_\text{pa}$ converges to neither zero nor one, but to a universal value, which depends only on the universality class, the aspect ratio of the torus, and the symmetry used for the twist.
These values can be obtained from conformal field theory (CFT) on a torus with modular parameter $\tau$ ($\tau = i$ for the $L\times L$ torus considered here).
Twisting the boundary condition by the $\mathbb{Z}_2$ symmetry in the space direction gives the partition function of the twisted sector, while twisting it in the time direction corresponds to the insertion of the $\mathbb{Z}_2$ operator into the trace.
On the critical line of the AT universality class, the universal values vary continuously along the line, since the twisted partition functions of the $c=1$ CFT depend on the compactification radius (see Appendix~\ref{appendix:cft}).

%%%%%%%%%%%%%%%%%%%%%%%%%%%%%%%%%%%%%%%%%%%%%%%%%%%%%%%%%%%%%%%%%
\subsection{Partition function on the Klein bottle}
\label{sec:klein-bottle}

On the critical line of the AT universality class for $g\geq g^*$, the partition-function ratios on a torus vary smoothly with the compactification radius and show no special feature at the critical endpoint $g=g^*$, where the universality class is that of the 4-state Potts model.
To examine this Potts point, we consider the partition functions on the Klein bottle.

The Potts point is characterized by its enhanced symmetry.
The continuous phase transition of the AT universality class has the symmetry of the dihedral group $D_4$, while the Potts point has the higher symmetry of the permutation group $S_4$.
These symmetries are associated with the four equivalent states of the stripe-ordered phase, which we denote by
$H_\pm$ and $V_\pm$ ($H_+$ is shown in Fig.~\ref{fig:stripe_lattice}).
Here, $H$ and $V$ represent the direction of the stripes and $\pm$ the sign of the spin at the origin.
While $S_4$ includes all the permutations of the four states, $D_4$ consists only of those mapping each of the pairs $\{H_+, H_-\}$ and $\{V_+, V_-\}$ onto either itself or the other pair.

To detect the enhancement from $D_4$ to $S_4$, we consider two elements of $D_4$, $d_1 = (H_+\,H_-)(V_+\,V_-)$ and $d_2 = (H_+\,V_+)(H_-\,V_-)$.
The former corresponds to the global spin flip, while the latter maps $H_{\pm}$ to $V_{\pm}$ with the same sign.
Both are double transpositions, i.e., products of two disjoint transpositions, and thus conjugate in $S_4$.
In $D_4$, however, they belong to different conjugacy classes, since $d_1$ maps each of the pairs onto itself while $d_2$ exchanges the two pairs.
Therefore, the partition functions with the boundary conditions twisted by $d_1$ and $d_2$ are expected to be different on the AT critical line except at the Potts point.

In the lattice model, however, the two elements are realized differently.
While $d_1$ is an internal symmetry of the model, $d_2$ is not: it is realized only in combination with a spatial operation.
Indeed, the reflection about a diagonal of the square lattice exchanges $H_\pm$ and $V_\pm$, and thus acts on the stripe states as $d_2$.
We therefore consider the partition functions on the Klein bottles $\mathcal{K}_A$ and $\mathcal{K}_D$, where the boundary condition in one direction is twisted by the reflection about a lattice axis ($x \rightarrow -x$) and a diagonal ($x \leftrightarrow y$), respectively.
As for the torus, we specify the boundary conditions for the spins by subscripts, $Z^{\mathcal{K}}_{bb'}$ with $b, b' \in \{\text{p}, \text{a}\}$, where $b$ and $b'$ refer to the periodic direction and the direction twisted by the reflection, respectively.
Then, $Z^{\mathcal{K}_D}_\text{pp}$ is twisted by $d_2$.
Since the reflection about a lattice axis leaves all the stripe states unchanged, the twist of $Z^{\mathcal{K}_A}_\text{pa}$ acts on them only through the antiperiodic boundary condition, i.e., as the global spin flip $d_1$.

\begin{figure}[t]
  \centering
  \includegraphics{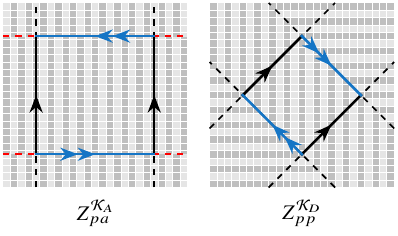}
  \caption{Schematic picture of the Klein-bottle partition functions $Z^{\mathcal{K}_A}_\text{pa}$ and $Z^{\mathcal{K}_D}_\text{pp}$, which are twisted by the reflection about a lattice axis and a diagonal, respectively.
  The arrow indicates how the opposite edges of the square are identified to form a Klein bottle.
  The direction of the spins is indicated by shades of gray, and the inside of the region is set to the $H_+$ stripe state.
  The horizontal red dashed line in $Z^{\mathcal{K}_A}_\text{pa}$ represents the antiperiodic boundary condition, which flips the spins across it.
  }
  \label{fig:stripe_klein}
\end{figure}

Figure~\ref{fig:stripe_klein} shows schematic pictures of the Klein-bottle partition functions $Z^{\mathcal{K}_A}_\text{pa}$ and $Z^{\mathcal{K}_D}_\text{pp}$.
The antiperiodic boundary condition in $Z^{\mathcal{K}_A}_\text{pa}$ forces a domain wall between a stripe state and its flipped state, e.g., between $H_+$ and $H_-$, which is equivalent to the action of $d_1$.
On the other hand, $Z^{\mathcal{K}_D}_\text{pp}$ is twisted by the reflection about a diagonal, which maps $H_+$ to $V_+$ and thus forces a domain wall corresponding to $d_2$.

In the continuum limit, the two reflections are related to each other by a spatial rotation, and both $d_1$ and $d_2$ become internal symmetries.
In CFT, the partition function on the Klein bottle twisted by an internal symmetry operation $g$ combined with the reflection is written as
\begin{equation}
  \mathcal{Z}[\Omega g]
  = \mathrm{Tr}\left(g\,\Omega\,e^{-\ell \mathcal{H}}\right),
  \label{eq:ZK_CFT}
\end{equation}
where $\mathcal{H}$ is the Hamiltonian of the CFT on a circle of circumference $L$, $\ell$ is the length of the Klein bottle in the direction twisted by the reflection, and $\Omega$ reverses the orientation of the circle.
Therefore, $Z^{\mathcal{K}_A}_\text{pa}$ and $Z^{\mathcal{K}_D}_\text{pp}$ correspond to $\mathcal{Z}[\Omega d_1]$ and $\mathcal{Z}[\Omega d_2]$, respectively, up to a non-universal factor due to the bulk free energy.
Since $d_1$ and $d_2$ are conjugate in $S_4$, $\mathcal{Z}[\Omega d_1] = \mathcal{Z}[\Omega d_2]$ holds at the Potts point, while they differ elsewhere on the critical line of the AT universality class.
Note that the same trace without $\Omega$, $\mathcal{Z}[g]$, gives the partition function on the torus twisted by $g$ in the direction of $\ell$ (see Appendix~\ref{appendix:cft}).

\begin{figure}[t]
  \centering
  \includegraphics{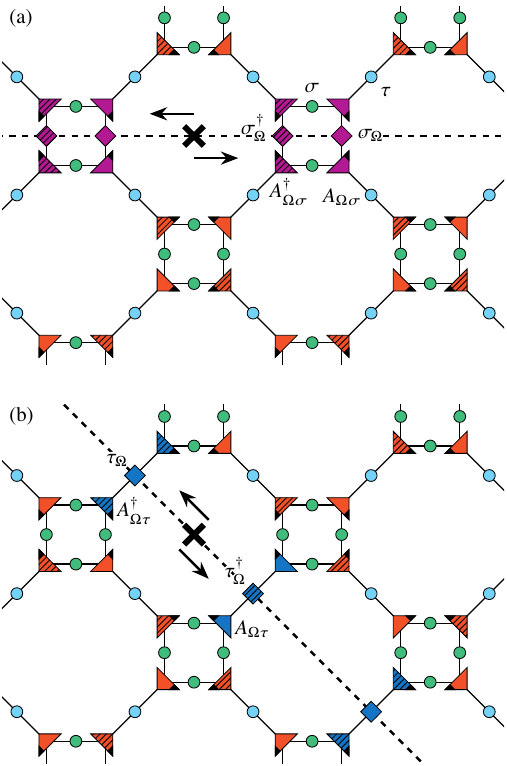}  
  \caption{Boundary tensors for the Klein bottle.
  (a) The boundary tensors $(\sigma_\Omega, A_{\Omega\sigma})$ on the reflection line through the $\sigma$ bonds.
  (b) The boundary tensors $(\tau_\Omega, A_{\Omega\tau})$ on the reflection line through the $\tau$ bonds.
  The dashed lines represent the reflection line, and the arrows indicate that the orientation is reversed across them.
  The cross marks the center of the reflection.
  The boundary tensors are shown in purple and blue, and the bulk tensors are the same as Fig.~\ref{fig:stripe_lattice}.}
  \label{fig:klein_tn}
\end{figure}
We compute the Klein-bottle partition functions by extending the BWTRG algorithm.
To glue the two ends of the system through the reflection, we introduce boundary tensors along the reflection line.
We place the reflection line perpendicular to the bonds so that it passes through the bonds, as shown in Fig.~\ref{fig:klein_tn}.
The bond matrices on the reflection line and the three-index tensors connected to them are distinguished from the bulk tensors and referred to as the boundary tensors.
Corresponding to the two types of reflection, about a lattice axis and about a diagonal, there are two types of boundary tensors.
Since the lattice is rotated by $45$ degrees in each step of BWTRG, a reflection about a lattice axis is transformed into a reflection about a diagonal and vice versa.
Thus, the two types of boundary tensors transform into each other in each step, and $Z^{\mathcal{K}_A}$ and $Z^{\mathcal{K}_D}$ are obtained at alternate steps.
Details of the update procedure and the contraction for the partition function are described in Appendix~\ref{appendix:klein}.

%%%%%%%%%%%%%%%%%%%%%%%%%%%%%%%%%%%%%%%%%%%%%%%%%%%%%%%%%%%%%%%%%
\section{Results}
\label{sec:results}

We perform the BWTRG calculations with the bond dimension up to $\chi=192$.
We always set the hyperparameter of BWTRG as $k=1/2$, which is considered to be optimal~\cite{adachi2022bondweighted}.

\subsection{Phase diagram}
\label{sec:results_tc}

The transition point is determined by the bisection method in $\beta$, where each temperature is classified into the disordered or ordered phase according to whether $R_\text{pa}$ converges to one or zero as the system size $L$ increases.
To reduce the computational cost, $R_\text{pa}$ in the bisection is computed from the smaller cluster [Eq.~\eqref{eq:trace_4a} in Appendix~\ref{appendix:bulk_tensor}].
We confirmed that the larger cluster [Eq.~\eqref{eq:trace_16a}] gives the same classification.
The bisection is repeated until the interval of $\beta$ becomes smaller than $10^{-9}$.

\begin{figure}[t]
  \centering
  \includegraphics{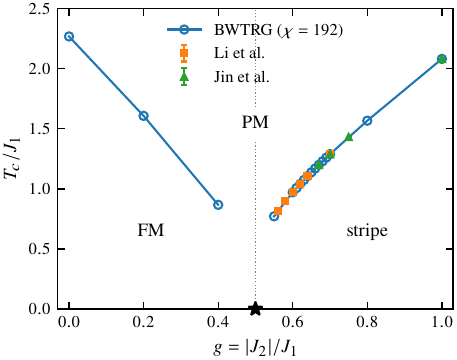}
  \caption{Phase diagram of the frustrated Ising model on the square lattice obtained by BWTRG with $\chi=192$.
  The finite bond-dimension effects are smaller than the symbol size.
  The Monte Carlo (MC) and transfer matrix (TM) results are taken from Refs.~\cite{li2024comprehensive} and \cite{jin2013phase}, respectively.
  }
  \label{fig:phase_diagram}
\end{figure}
Figure~\ref{fig:phase_diagram} shows the phase diagram obtained by BWTRG with $\chi=192$, and the values of $T_\text{c}$ are listed in Table~\ref{tab:tc} in Appendix~\ref{appendix:tc}.
At $g=0$, $T_\text{c}$ agrees with the exact value $2/\ln(1+\sqrt{2})$ within the error.
Our results also agree with those of the MC~\cite{li2024comprehensive} and transfer matrix (TM) methods~\cite{jin2013phase} within the error bars, while the accuracy is improved by several orders of magnitude.

\subsection{Partition function ratios on a torus}
\label{sec:results_torus}

\begin{figure}[t]
  \centering
  \includegraphics{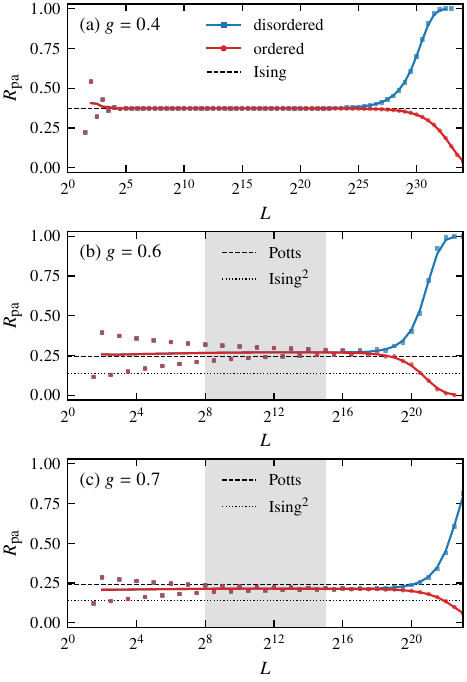}
  \caption{Size dependence of the partition-function ratio $R_\text{pa}$ on a torus at (a) $g=0.4$, (b) $g=0.6$, and (c) $g=0.7$ calculated by BWTRG with $\chi=192$.
  The points are the raw data of BWTRG with $\chi=192$ and the solid lines are the average over neighboring steps defined in Eq.~\eqref{eq:rpa_average} to suppress the oscillation.
  The dashed line in (a) is the universal value predicted by the Ising CFT.
  The dashed (dotted) line in (b, c) is the universal value for the 4-state Potts ($\text{Ising}^2$) CFT.
  The shaded area indicates the range for estimating the plateau value shown in Fig.~\ref{fig:rpa_plateau}.
  }
  \label{fig:rpa_16a}
\end{figure}
Figure~\ref{fig:rpa_16a} shows the size dependence of the partition-function ratio $R_\text{pa}$ at temperatures on both sides of $T_\text{c}$ on a torus at (a) $g=0.4$, (b) $g=0.6$ and (c) $g=0.7$.
Hereafter, the partition-function ratios are computed from the larger cluster [Eq.~\eqref{eq:trace_16a}].

The raw data of $R_\text{pa}$ oscillate between even and odd BWTRG steps, which is due to the $45^\circ$ rotation of the lattice in each step.
The cluster in Eq.~\eqref{eq:trace_16a} corresponds to the $L\times L$ torus aligned with the lattice axes at odd steps ($L=2^k$) and to that rotated by $45^\circ$ at even steps, and the former gives larger values of $R_\text{pa}$.
For $g<1/2$, the anisotropy distinguishing the two orientations comes only from the spin-4 lattice perturbation with the scaling dimension $\Delta=4$.
Its contribution is proportional to $L^{2-\Delta}$, and thus the amplitude rapidly decreases by a factor of $(\sqrt{2})^{-2}=1/2$ in each BWTRG step.
For $g>1/2$, in contrast, the domain walls of the stripe order are strongly anisotropic, and the corresponding spin-2 perturbation $\partial_x\sigma\,\partial_x\tau-\partial_y\sigma\,\partial_y\tau$ in the AT description has the scaling dimension $\Delta'=2+x_P$, close to marginal.
Here, $x_P$ is the scaling dimension of the polarization operator $\sigma\tau$, which takes $x_P=1/8$ at the 4-state Potts point and increases toward $1/4$ along the AT critical line as $g$ increases.
Since this perturbation is odd under $\tau\to-\tau$, its first-order contribution vanishes, and the second-order one
scales as $L^{2(2-\Delta')}$.
The amplitude therefore decreases only by a factor of $2^{-x_P}$ per step, which is consistent with the observed slow decay.
Since $x_P$ increases with $g$, the oscillation decays faster for larger $g$, as seen by comparing Figs.~\ref{fig:rpa_16a}(b) and (c).
To suppress the oscillation, we use the average over neighboring steps,
\begin{equation}
  \bar{R}(L) \equiv \frac{1}{4}\left[
    R\left(L/\sqrt{2}\right) + 2 R\left(L\right) + R\left(\sqrt{2}L\right)
    \right],
  \label{eq:rpa_average}
\end{equation}
which is shown by the solid lines in Fig.~\ref{fig:rpa_16a}.
If the oscillating part behaves as $(-1)^n A \lambda^n$ with the step $n$, the average in Eq.~\eqref{eq:rpa_average} reduces its amplitude by a factor of $(1-\lambda)^2/(4\lambda)$.

Before converging to one or zero, $R_\text{pa}$ stays at a constant value over a wide range of $L$, which indicates that $R_\text{pa}$ takes a universal value at criticality similarly to the Binder ratio.
For $g=0.4$, the plateau value agrees with the universal value of the Ising CFT.
We confirmed that the plateau values for $g=0$ and $0.2$ also agree with the Ising CFT prediction, which is consistent with the fact that the transition between the ferromagnetic and paramagnetic phases belongs to the Ising universality class.
For $g=0.6$ and $0.7$, the plateau values are different from the Ising value, as expected for the stripe transition.
The plateau value for $g=0.7$ is between the universal values of the 4-state Potts and $\text{Ising}^2$ CFTs, which is consistent with the fact that $g=0.7$ is on the AT critical line.
On the other hand, the plateau value for $g=0.6$ exceeds the universal value at the 4-state Potts point, which suggests that $g=0.6$ is located on the first-order side of the Potts point.
For $1/2<g<g^*$, where the transition is expected to be weakly first order, the plateau is presumably a pseudo-critical one: the renormalization-group flow slowly passes near the extension of the fixed line corresponding to the AT critical line beyond the Potts point, where the operator that is marginal at the Potts point becomes weakly relevant.
In this case, the plateau value for $g<g^*$ is not universal, but it still reflects how close the flow passes to the Potts point.

\begin{figure}[t]
  \centering
  \includegraphics{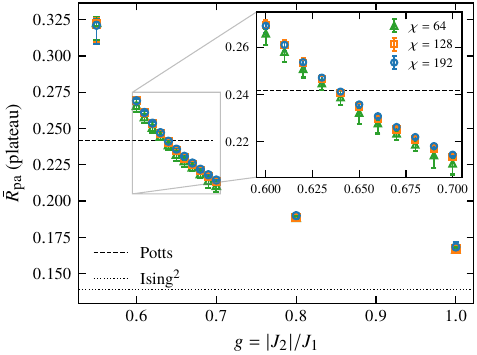}
  \caption{Plateau values of the partition-function ratio $\bar{R}_\text{pa}$ for $g>1/2$ obtained with $\chi=64$, $128$, and $192$.
  The points are the averages of $\bar{R}_\text{pa}(L)$ in the range $2^8\leq L\leq 2^{15}$ ($2^8\leq L\leq 2^{13}$ for $g=0.55$), and the error bars indicate its maximum and minimum values in the range.
  For $\chi=64$, the range is restricted to $2^8\leq L\leq 2^{12}$ for $g=0.8$ and $2^8\leq L\leq 2^{10}$ for $g=1.0$.
  The dashed (dotted) line is the universal value for the 4-state Potts ($\text{Ising}^2$) CFT.
  The inset magnifies the region $0.6\leq g\leq 0.7$ around the crossing with the Potts value.}
  \label{fig:rpa_plateau}
\end{figure}
Figure~\ref{fig:rpa_plateau} shows the plateau values of $R_\text{pa}$ for $g>1/2$ obtained with $\chi=64$, $128$, and $192$.
We estimate the plateau value by averaging $\bar{R}_\text{pa}(L)$ over the range $2^8\leq L\leq 2^{15}$, where the difference of $R_\text{pa}$ between the two temperatures of the final bisection interval is less than $10^{-3}$.
For $g=0.55$, where two temperatures separate at smaller $L$, we use the range $2^8\leq L\leq 2^{13}$.
For $\chi=64$, the range is restricted to $2^8\leq L\leq 2^{12}$ for $g=0.8$ and $2^8\leq L\leq 2^{10}$ for $g=1.0$, where $\bar{R}_\text{pa}$ deviates from the plateau at larger $L$ due to the finite bond dimension.

The plateau value decreases monotonically as $g$ increases, toward the universal value of the $\text{Ising}^2$ CFT expected for $g\to\infty$.
Since it changes smoothly across $g^*$, the Potts point can be located by its crossing with the universal value of the 4-state Potts CFT, which occurs between $g=0.63$ and $0.64$ and suggests $g^*\approx 0.64$.
The plateau values shift upward with increasing $\chi$ in the vicinity of the crossing, and the crossing point moves from $g=0.635$ for $\chi=64$ to $0.638$ for $\chi=128$ and $0.639$ for $\chi=192$.
Although the shift decreases with increasing $\chi$, the above estimate of $g^*$ is likely to be a lower bound of the location of the Potts point.

The torus ratios locate the Potts point only by matching their values with that of the Potts CFT, which is sensitive to the logarithmic corrections due to the marginal operator at the Potts point.
In the next subsection, we therefore examine the ratio of the Klein-bottle partition functions, which directly detects the enhanced $S_4$ symmetry.

\subsection{Partition function ratios on a Klein bottle}
\label{sec:results_klein}

We consider two ratios of the partition functions,
\begin{equation}
  R^{\mathcal{K}_A}_\text{pa} \equiv \frac{Z^{\mathcal{K}_A}_\text{pa}}{Z_\text{pp}}, \quad
  R^{\mathcal{K}_D}_\text{pp} \equiv \frac{Z^{\mathcal{K}_D}_\text{pp}}{Z_\text{pp}},
  \label{eq:klein_ratios}
\end{equation}
where the Klein bottles and the torus have the same size $L\times L$, so that the non-universal factor due to the bulk free energy cancels out.
As mentioned in Sec.~\ref{sec:klein-bottle}, on the AT critical line the two ratios are expected to coincide only at the Potts point, where $D_4$ is enhanced to $S_4$.
Therefore, the crossing point of the two ratios provides an estimate of the Potts point that is independent of the universal value of the 4-state Potts CFT.

Before discussing the stripe transition, we confirmed the validity of the computation of the Klein-bottle partition functions for $g<1/2$, where the transition belongs to the Ising universality class.
In this case, the reflection about a diagonal does not act on the order parameter, and $R^{\mathcal{K}_A}_\text{pa}$ and $R^{\mathcal{K}_D}_\text{pp}$ correspond to $\mathcal{Z}[\Omega d_1]/\mathcal{Z}[1]$ and $\mathcal{Z}[\Omega]/\mathcal{Z}[1]$ of the Ising CFT, respectively.
The plateau values for $g=0$, $0.2$, and $0.4$ agree with the universal values, $0.37404$ and $0.99882$, within $10^{-4}$.

\begin{figure}[t]
  \centering
  \includegraphics{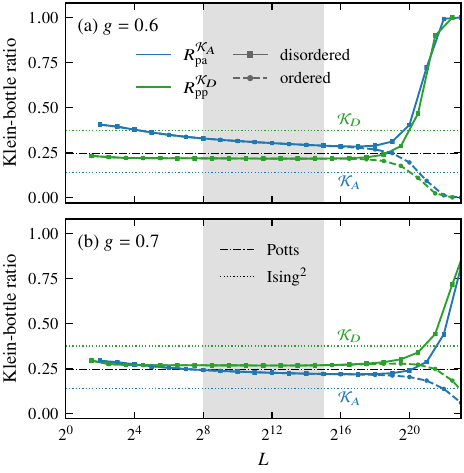}
  \caption{Klein-bottle partition-function ratios $R^{\mathcal{K}_A}_\text{pa}$ and $R^{\mathcal{K}_D}_\text{pp}$ at (a) $g=0.6$ and (b) $g=0.7$ obtained with $\chi=192$.
  The solid (dashed) lines with squares (circles) are the results at the temperature on the disordered (ordered) side of the final bisection interval.
  The dash-dotted line is the universal value of the 4-state Potts CFT, which is common to the two ratios.
  The dotted lines labeled $\mathcal{K}_A$ and $\mathcal{K}_D$ are the universal values of $R^{\mathcal{K}_A}_\text{pa}$ and $R^{\mathcal{K}_D}_\text{pp}$ for the $\text{Ising}^2$ CFT, respectively.
  The shaded area indicates the range for estimating the plateau values shown in Fig.~\ref{fig:klein_plateau}.
  }
  \label{fig:klein_16a}
\end{figure}
Figure~\ref{fig:klein_16a} shows the size dependence of the Klein-bottle partition-function ratios $R^{\mathcal{K}_A}_\text{pa}$ and $R^{\mathcal{K}_D}_\text{pp}$ at (a) $g=0.6$ and (b) $g=0.7$.
The partition functions on the Klein bottle, $Z^{\mathcal{K}_A}_\text{pa}$ and $Z^{\mathcal{K}_D}_\text{pp}$, are obtained at odd and even BWTRG steps, respectively.
Since each ratio is obtained only at every other step, it does not show the even-odd oscillation observed in $R_\text{pa}$.
As in the case of $R_\text{pa}$, both ratios converge to one in the disordered phase and to zero in the ordered phase, while they show plateaus at intermediate sizes.

The slow decrease of $R^{\mathcal{K}_A}_\text{pa}$ is observed for all $g$ in $0.6\leq g\leq 0.7$, while $R_\text{pa}$ and $R^{\mathcal{K}_D}_\text{pp}$ show much weaker size dependence in the same range.
This suggests that the finite-size corrections to $R^{\mathcal{K}_A}_\text{pa}$ decay slowly, possibly due to the marginal operator at the Potts point.
Since the decrease is monotonic, $R^{\mathcal{K}_A}_\text{pa}$ is expected to approach its fixed-point value from above.
Consequently, the average of $R^{\mathcal{K}_A}_\text{pa}$ over the range overestimates the universal value.

We therefore estimate the plateau values as follows.
For $R^{\mathcal{K}_D}_\text{pp}$, we take the average over $2^8\leq L\leq 2^{15}$ and the two temperatures of the final bisection interval, as in the case of $R_\text{pa}$.
For $R^{\mathcal{K}_A}_\text{pa}$, we use the value at $L=2^{15}$ on the disordered side, i.e., the minimum in the range, as an upper bound of the universal value.
This is justified by two observations:
the value on the disordered side is larger than that on the ordered side at all steps, so that the renormalization-group flow from the critical point lies between them;
and the value on the disordered side decreases monotonically before the two temperatures separate.
We also tried the extrapolation with a power-law correction, $a+bL^{-c}$, but it was not stable near the Potts point, which is consistent with logarithmic corrections due to the marginal operator.

\begin{figure}[t]
  \centering
  \includegraphics{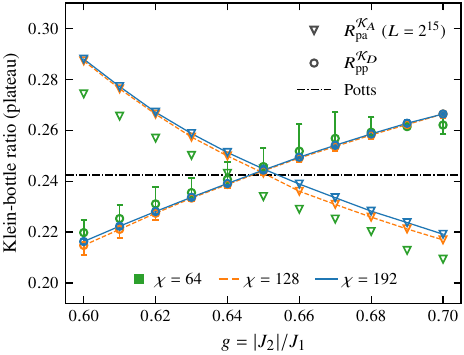}
  \caption{Plateau values of the Klein-bottle partition-function ratios $R^{\mathcal{K}_A}_\text{pa}$ and $R^{\mathcal{K}_D}_\text{pp}$ for $0.6\leq g \leq 0.7$ obtained with $\chi=64$, $128$, and $192$ (colors).
  For $R^{\mathcal{K}_A}_\text{pa}$, the value at $L=2^{15}$ in the disordered phase is plotted as an upper bound of the universal value.
  For $R^{\mathcal{K}_D}_\text{pp}$, the points are the average over $2^8\leq L\leq 2^{15}$, and the error bars indicate its maximum and minimum values in the range.
  The dash-dotted line is the universal value of the 4-state Potts CFT, which is common to the two ratios.
  }
  \label{fig:klein_plateau}
\end{figure}
Figure~\ref{fig:klein_plateau} shows the plateau values for $0.6\leq g\leq 0.7$.
$R^{\mathcal{K}_A}_\text{pa}$ decreases and $R^{\mathcal{K}_D}_\text{pp}$ increases with $g$, and they cross each other at $g=0.650$ for $\chi=192$.
Since $R^{\mathcal{K}_A}_\text{pa}$ in the figure is an upper bound, the true crossing point, i.e., the Potts point, is located at a smaller $g$.
Indeed, if the minimum is taken up to $L=2^{17}$ beyond the range, the crossing point moves to $g=0.647$.
Conversely, if the average over $2^8\leq L\leq 2^{15}$ is used for $R^{\mathcal{K}_A}_\text{pa}$ as for $R^{\mathcal{K}_D}_\text{pp}$, the crossing point is located at $g=0.660$, which shows that the slow decrease of $R^{\mathcal{K}_A}_\text{pa}$ shifts the estimate toward larger $g$.
The finite bond dimension, on the other hand, shifts the crossing point in the opposite direction: $R^{\mathcal{K}_A}_\text{pa}$ slightly increases with $\chi$, and the crossing point moves from $g=0.649$ for $\chi=128$ to $0.650$ for $\chi=192$.
This shift is smaller than that due to the finite size, and it decreases with increasing $\chi$.
We thus regard $g=0.650$ as an upper bound of $g^*$.
The plateau value of $R^{\mathcal{K}_D}_\text{pp}$, which shows almost no size dependence, crosses the universal value of the Potts CFT at $g=0.646$ for both $\chi=128$ and $192$.

The estimates of $g^*$ are summarized as follows.
The crossing of $\bar R_\text{pa}$ with the Potts value gives $g=0.639$, which is likely to be a lower bound as discussed in Sec.~\ref{sec:results_torus}, and the crossing of the two Klein-bottle ratios gives an upper bound $g=0.650$.
The estimate from $R^{\mathcal{K}_D}_\text{pp}$, $g=0.646$, lies between them.
We therefore conclude $0.639\lesssim g^*\lesssim 0.650$.

\begin{figure}[t]
  \centering
  \includegraphics{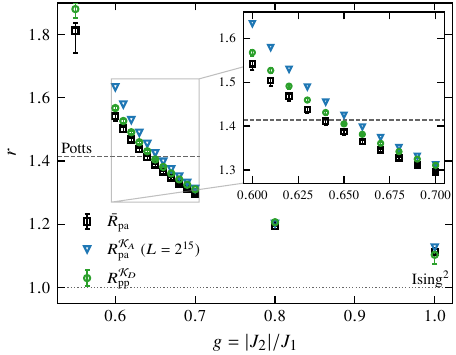}
  \caption{Compactification radius $r$ of the $\mathbb{Z}_2$ orbifold CFT estimated from the partition-function ratios for $\chi=192$.
  For each ratio, $r$ is determined so that the universal value given in Appendix~\ref{appendix:cft} coincides with the plateau value of $\bar{R}_\text{pa}$ or $R^{\mathcal{K}_D}_\text{pp}$, or with the value of $R^{\mathcal{K}_A}_\text{pa}$ at $L=2^{15}$ in the disordered phase, which gives an upper bound of $r$.
  The error bars correspond to the range of the plateau values shown in Figs.~\ref{fig:rpa_plateau} and \ref{fig:klein_plateau}.
  The dashed and dotted lines indicate the 4-state Potts ($r=\sqrt{2}$) and $\text{Ising}^2$ ($r=1$) points, respectively.
  The value from $R^{\mathcal{K}_A}_\text{pa}$ at $g=0.55$ ($r\approx2.1$) is out of the range.
  The inset magnifies the region $0.6\leq g\leq 0.7$.}
  \label{fig:r_plateau}
\end{figure}
To compare the three ratios on an equal footing, we convert each of them to the compactification radius $r$ of the $\mathbb{Z}_2$ orbifold CFT, using the $r$ dependence of the universal values shown in Fig.~\ref{fig:cft_ratios} in Appendix~\ref{appendix:cft}.
Figure~\ref{fig:r_plateau} shows $r$ estimated from the plateau values of $\bar{R}_\text{pa}$ and $R^{\mathcal{K}_D}_\text{pp}$ and from the value of $R^{\mathcal{K}_A}_\text{pa}$ at $L=2^{15}$, which gives an upper bound of $r$.
The three estimates of $r$ decrease monotonically with $g$, and their crossings with $r=\sqrt{2}$ are equivalent to the crossings of the ratios with the universal values of the 4-state Potts CFT. The three estimates agree with each other closely, which supports that the plateaus of the three ratios are described by the same $\mathbb{Z}_2$ orbifold CFT.
The radius $r$ is directly related to the critical exponents on the AT critical line.
The scaling dimensions of the energy and polarization operators are given by $x_\varepsilon=1/r^2$ and $x_P=1/(4r^2)$, and thus the correlation-length exponent is $\nu=1/(2-x_\varepsilon)=r^2/(2r^2-1)$, which varies from $\nu=1$ at the $\text{Ising}^2$ point to $2/3$ at the Potts point, whereas the scaling dimension $x_\sigma=1/8$ of the magnetization is independent of $r$~\cite{ginsparg1988applied}.
For $g<g^*$, the estimated $r$ exceeds $\sqrt{2}$, which corresponds to the pseudo-critical plateau near the extension of the AT critical line beyond the Potts point.

%%%%%%%%%%%%%%%%%%%%%%%%%%%%%%%%%%%%%%%%%%%%%%%%%%%%%%%%%%%%%%%%%
\section{Summary and Discussion}
\label{sec:summary}

In this paper, we have investigated the phase transition of the frustrated $J_1$--$J_2$ Ising model on the square lattice using the BWTRG method.
Using the block structure of the $\mathbb{Z}_2$ symmetric TN, we computed the partition functions on a torus with different boundary conditions, $Z_\text{pp}$ and $Z_\text{pa}$, separately.
We also introduced the boundary tensors on the reflection line, which allow us to compute the partition functions on the Klein bottle with a finite aspect ratio, including those twisted by internal symmetry operations.

The transition temperatures determined from $R_\text{pa}$ agree with the previous MC and TM results, while the accuracy is improved by several orders of magnitude.
At the transition, $R_\text{pa}$ shows a plateau over a wide range of $L$.
For $g<1/2$, the plateau value agrees with the universal value of the Ising CFT.
For $g>1/2$, the plateau value decreases monotonically with $g$ toward the universal value of the $\text{Ising}^2$ CFT, and it lies between the universal values of the 4-state Potts and $\text{Ising}^2$ CFTs on the AT critical line.
The plateau also appears for $1/2<g<g^*$, where the transition is weakly first order, reflecting the slow renormalization-group flow near the Potts point.
Since the plateau value changes smoothly across $g^*$, its crossing with the universal value of the 4-state Potts CFT gives an estimate $g=0.639$ for $\chi=192$.
This estimate is likely to be a lower bound of $g^*$, since the plateau values near the crossing increase with $\chi$.

We also examined the ratios of the Klein-bottle partition functions, $R^{\mathcal{K}_A}_\text{pa}$ and $R^{\mathcal{K}_D}_\text{pp}$, which are twisted by the global spin flip $d_1$ and by the exchange $d_2$ of the horizontal and vertical stripe states, respectively.
Since $d_1$ and $d_2$ are conjugate in $S_4$ but not in $D_4$, the two ratios coincide only at the Potts point, which allows us to locate it without relying on the universal value of the 4-state Potts CFT.
$R^{\mathcal{K}_A}_\text{pa}$ decreases slowly and monotonically with $L$, and we used its value at $L=2^{15}$ as an upper bound on the universal value.
The crossing of the two ratios thus gives an upper bound $g^*\lesssim 0.650$.
On the other hand, $R^{\mathcal{K}_D}_\text{pp}$ shows almost no size dependence, and its plateau value crosses the universal value of the 4-state Potts CFT at $g=0.646$.
We also note that $R^{\mathcal{K}_A}_\text{pa}>R^{\mathcal{K}_D}_\text{pp}$ for $g<g^*$, which corresponds to the extension of the AT critical line beyond the Potts point ($r>\sqrt{2}$) and is consistent with the pseudo-critical plateau of $R_\text{pa}$.

In conclusion, we estimate the location of the Potts point as $0.639\lesssim g^*\lesssim 0.650$, which is supported by the three estimates from the ratios on the torus and the Klein bottle.
This is smaller than the previous MC estimates, $g^*=0.67$~\cite{jin2012ashkinteller,kalz2012location,jin2013phase} and $0.696(10)$~\cite{sun2026relevantmarginal}, which were obtained from the system sizes of $L\sim10^2$--$10^3$.
Near $g^*$, the finite-size corrections decay slowly due to the marginal operator at the Potts point, and the estimates from smaller sizes tend to be shifted toward larger $g$.
Indeed, if we use the average of $R^{\mathcal{K}_A}_\text{pa}$ over $2^8\leq L\leq 2^{15}$ instead of the value at $L=2^{15}$, the crossing of the two Klein-bottle ratios moves to $g\approx 0.66$.
In addition, the even-odd oscillation of $R_\text{pa}$ in Fig.~\ref{fig:rpa_16a} indicates that the torus aligned with the lattice axes, which is used in conventional MC simulations, has a positive correction due to the anisotropy of the stripe domain walls, which decays only as $L^{-2x_P}\approx L^{-1/4}$ near the Potts point.
If only the aligned tori are used, the crossing of $R_\text{pa}$ with the Potts value is located at $g\approx0.70$ for $L=128$ and $0.675$ for $L=1024$, comparable to the previous MC estimates, and it is still $0.652$ for $L=2^{15}$.
The system sizes up to $L=2^{15}\approx 3\times10^4$ accessible by BWTRG are thus essential for locating the Potts point.
On the other hand, a recent MC study suggested $g^*\lesssim0.62$, regarding the transitions for $g\geq0.58$ as continuous~\cite{li2024comprehensive}.
The values of $\nu$ reported there, which are smaller than the Potts value $2/3$ for $g\leq0.62$, agree with those obtained from our estimates of $r$ through $\nu=r^2/(2r^2-1)$ at $g=0.60$ and $0.62$, where $r>\sqrt{2}$.
This suggests that the apparent continuous transitions for $g\lesssim0.62$ reflect the pseudo-critical behavior below the Potts point.

We note that these estimates are based on the assumption that the continuous transition for $g>g^*$ belongs to the AT universality class and that the AT critical line terminates at the 4-state Potts point.
Under this assumption, the universal values of the ratios are given by the $\mathbb{Z}_2$ orbifold CFT, and the Potts point is identified as the point where the symmetry is enhanced from $D_4$ to $S_4$.
In addition, the estimate from $\bar{R}_\text{pa}$ assumes that the plateau value, which is pseudo-critical for $g<g^*$, changes smoothly across $g^*$, and it is regarded as a lower bound since the crossing point moves toward larger $g$ with increasing $\chi$.
The upper bound from the Klein-bottle ratios relies on the monotonic approach of $R^{\mathcal{K}_A}_\text{pa}$ to its universal value from above, and the estimate from $R^{\mathcal{K}_D}_\text{pp}$ assumes that it does not drift beyond the accessible system sizes.
The effect of the finite bond dimension on the Klein-bottle estimates is small: the crossing points change by at most $10^{-3}$ between $\chi=128$ and $192$.

The present results also provide some information on the scenarios different from the above assumption.
The three ratios give consistent values of $r$ for $0.6\leq g\leq 0.7$, which are also consistent with $R_\text{aa}$ as discussed below, indicating that the plateaus in this range are described by the $\mathbb{Z}_2$ orbifold CFT with $c=1$ rather than by the tricritical Ising CFT with $c=7/10$ proposed in Ref.~\cite{li2021tensor}.
If the transition were weakly first order also for $g>g^*$ as argued in Ref.~\cite{gangat2024weak}, the plateaus would be pseudo-critical, as in the region $1/2<g<g^*$.
However, the estimated $r$ is smaller than $\sqrt{2}$ for $g>g^*$, and the sign of $R^{\mathcal{K}_A}_\text{pa}-R^{\mathcal{K}_D}_\text{pp}$ is reversed across $g^*$, in agreement with the AT critical line inside the region $1<r<\sqrt{2}$.
Nevertheless, the present data cannot exclude an extremely weak first-order transition whose correlation length exceeds $L=2^{15}$.

Other partition-function ratios can also be computed within the same framework.
A quantity commonly used in TRG studies is the Gu-Wen ratio $X_1=Z(L\times L)^2/Z(2L\times L)$~\cite{gu2009tensorentanglementfiltering,morita2026tensor}, which counts the degeneracy of the fixed-point tensor.
We note that $X_1$ constructed from $Z^\text{TN}$ in Eq.~\eqref{eq:ZTN}, which is the sum over the four boundary conditions, takes the value $2$ in both the disordered and stripe phases and cannot distinguish them.
In contrast, $X_1^\text{pp}$ constructed from $Z_\text{pp}$ takes $1$, $2$, and $4$ in the disordered, ferromagnetic, and stripe phases, respectively.
On the AT critical line, the universal values of $X_1^\text{pp}$ and $R_\text{aa}\equiv Z_\text{aa}/Z_\text{pp}$ are also functions of the compactification radius $r$, which can be obtained from the $\mathbb{Z}_2$ orbifold CFT in the same way as in Appendix~\ref{appendix:cft}.
The values of $r$ obtained from the plateau values of $R_\text{aa}$ agree with those from $\bar{R}_\text{pa}$ in Fig.~\ref{fig:r_plateau} within $0.01$ for $g\geq0.6$ (not shown), which further supports that the plateaus are described by the $\mathbb{Z}_2$ orbifold CFT.
The value of $r$ from $X_1^\text{pp}$ is systematically smaller by about $0.02$ and crosses $r=\sqrt{2}$ at $g=0.628$, while that from $R_\text{aa}$ crosses it at $g=0.637$, close to $g=0.639$ from $\bar{R}_\text{pa}$.
This is possibly because $X_1^\text{pp}$ involves the partition function with a different aspect ratio, which has different finite-size corrections, and because $X_1^\text{pp}$ is less sensitive to $r$ than $R_\text{pa}$ and $R_\text{aa}$.

Finally, we comment on possible extensions of the present method.
The boundary tensors on the reflection line can be introduced in other TRG-type algorithms and models, although the construction of the initial boundary tensors should be examined for each model.
Since the Klein-bottle partition functions twisted by internal symmetry operations directly detect the conjugacy classes of the symmetry group, they provide a general tool to identify emergent symmetries at critical points, such as the enhancement from $D_4$ to $S_4$ discussed here.
It would also be interesting to extend the method to the Klein bottle with a large aspect ratio, which gives the universal Klein-bottle entropy~\cite{tu2017universal}, and to the real projective plane~\cite{shimizu2026tensor}.
As for the accuracy, BWTRG does not remove the short-range correlations in the loops of the network, which limits the accuracy at large system sizes.
Combining the present method with renormalization schemes that include entanglement filtering, such as tensor network renormalization (TNR)~\cite{evenbly2015tensor}, loop-TNR~\cite{yang2017loop}, graph-independent local truncations (GILT)~\cite{hauru2018renormalization}, and entanglement branching operators~\cite{harada2018entanglement}, is expected to further improve the accuracy, especially near the Potts point, where the slow finite-size corrections require large system sizes.

%%%%%%%%%%%%%%%%%%%%%%%%%%%%%%%%%%%%%%%%%%%%%%%%%%%%%%%%%%%%%%%%%
\acknowledgments
S.M. is supported by the Center of Innovations for Sustainable Quantum AI (JST Grant Number JPMJPF2221).
The computation in this work has been done using the facilities of the Supercomputer Center, the Institute for Solid State Physics, the University of Tokyo (ISSPkyodo-SC-2026-Ba-0045).

\subsection*{DATA AVAILABILITY STATEMENT}
The data that support the findings of this article are available from the author upon reasonable request.

\subsection*{AI Use Disclosure}
In the preparation of this work, the author used Claude Opus 5.5 (Anthropic)
to assist in writing the codes for the simulations and the data analysis,
in checking the derivations of the CFT partition functions,
and in proofreading and editing the manuscript.
All the codes, results, and text were reviewed and verified by the author,
who takes full responsibility for the content of this article.

%%%%%%%%%%%%%%%%%%%%%%%%%%%%%%%%%%%%%%%%%%%%%%%%%%%%%%%%%%%%%%%%%
\appendix
\section{Details of BWTRG}
\label{appendix:bwtrg}

In this appendix, we describe the details of the BWTRG algorithm used in this paper.

\subsection{Update of the bulk tensors}
\label{appendix:bulk_tensor}

We consider the triad TN with the inner weight $\sigma$, the outer weight $\tau$, and the isometry $A$ shown in Fig.~\ref{fig:stripe_lattice}.
Following Ref.~\cite{lyu2026latticea}, we update the triad TN by the eigenvalue decomposition, which preserves the $90^\circ$ lattice-rotational symmetry.
This is in contrast to the singular value decomposition used in the original BWTRG~\cite{adachi2022bondweighted,morita2025multiimpurity}.
Since the eigenvalue can be negative, the signs remain in the bond weights rather than being absorbed into the isometries, which allows us to handle bond weights that can take negative values.

The outer weight $\tilde{\tau}$ and the isometry $\tilde{A}$ in the next step are obtained from the truncated eigenvalue decomposition of the plaquette on the left-hand side of the following equation:
\begin{equation}
  \includetikz{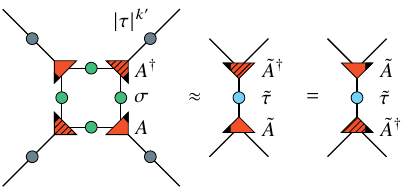},
  \label{eq:plaquette_decomp}
\end{equation}
where $k'\equiv (1-k)/2$ with the hyperparameter $k$ of BWTRG.
The plaquette can be reshaped into a Hermitian matrix between the upper and lower pairs of the external legs.
Its eigenvalue decomposition gives $\tilde{\tau}$ as the diagonal matrix of the real eigenvalues and $\tilde{A}$ as the isometry of the corresponding eigenvectors, where we retain up to the $\chi$ largest eigenvalues in absolute value.

The inner weight $\tilde{\sigma}$ in the next step is obtained from the outer weight $\tau$.
In the standard BWTRG, where the bond weights are positive, it is defined as $\tilde{\sigma}=\tau^k$.
Since $\tau$ can take negative values in our case, we define $\tilde{\sigma}$ to preserve the sign of $\tau$:
\begin{equation}
  \tilde{\sigma}
  = \operatorname{sgn} (\tau) \left|\tau\right|^{k}
  = \left|\tau\right|^{-k'} \tau \left|\tau\right|^{-k'}.
\end{equation}
The last expression shows that $|\tau|^{k'}\tilde{\sigma}|\tau|^{k'}=\tau$, i.e., the factors $|\tau|^{k'}$ attached to the plaquettes in Eq.~\eqref{eq:plaquette_decomp} and $\tilde{\sigma}$ between them reproduce the original bond weight $\tau$.

To make the construction applicable to symmetric tensors with conserved quantities, the direction in which the conserved quantity flows through each index must be specified.
We assign the directions of the indices of $A$ as follows:
\begin{equation}
  \includetikz{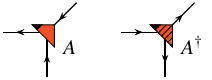}.
  \label{eq:A_flow}
\end{equation}
Then, the directions of all the indices in the BWTRG update, including those in Eq.~\eqref{eq:plaquette_decomp}, are determined consistently.
In this paper, the tensors are real and $\mathbb{Z}_2$ symmetric, for which the directions of the indices are irrelevant.
Therefore, we have $A^\dagger=A$, and the distinction between them is not necessary in practice.

The partition function on a torus can also be calculated by contracting the tensors in the unit cell:
\begin{equation}
  Z^\text{TN} = \includetikz{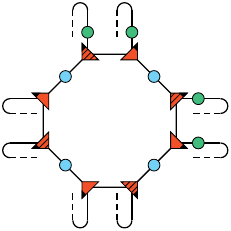},
  \label{eq:trace_8a}
\end{equation}
which requires the $O(\chi^6)$ computational cost, the same as Eq.~\eqref{eq:trace_16a}.
To further reduce the computational cost, the partition function for a cluster smaller than the unit cell by a factor of $\sqrt{2}$ can be calculated with the $O(\chi^5)$ computational cost as
\begin{equation}
  Z^\text{TN} = \includetikz{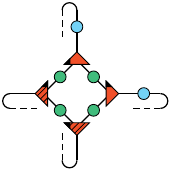},
  \label{eq:trace_4a}
\end{equation}
where we utilize the relation in Eq.~\eqref{eq:plaquette_decomp} to restore the diagonal translational symmetry of Fig.~\ref{fig:stripe_lattice}.

For the same system size, Eq.~\eqref{eq:trace_16a} requires one and two fewer BWTRG steps than Eqs.~\eqref{eq:trace_8a} and \eqref{eq:trace_4a}, respectively, and is thus more accurate.
We therefore use Eq.~\eqref{eq:trace_16a} to analyze the universal behavior of the partition-function ratios.
In the bisection for the transition temperature, we use Eq.~\eqref{eq:trace_4a} to reduce the computational cost, since it gives the same phase for each temperature as Eqs.~\eqref{eq:trace_16a} and \eqref{eq:trace_8a}.

\subsection{Update of the Klein-bottle boundary tensors}
\label{appendix:klein}

\begin{figure}[t]
  \centering
  \includegraphics{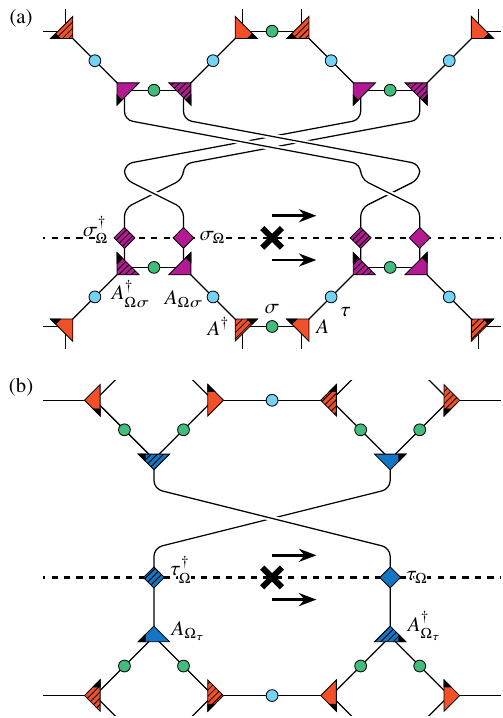}
  \caption{Update of the boundary tensors from $(\sigma_\Omega, A_{\Omega \sigma})$ to $(\tau_\Omega, A_{\Omega \tau})$ in a BWTRG step.
  (a) The tensor network around the reflection line before the update and (b) that after the update.
  Unlike Fig.~\ref{fig:klein_tn}, the network above the reflection line is drawn in the same orientation as that below it, as indicated by the arrows.
  Instead, the bonds crossing the reflection line connect the tensors at the mirror-image positions with respect to the center of the reflection (cross).
  The dashed lines, the colors, and the symbols are the same as in Fig.~\ref{fig:klein_tn}.
  }
  \label{fig:klein_update}
\end{figure}

In this subsection, we describe how the boundary tensors introduced in Fig.~\ref{fig:klein_tn} are updated together with the bulk tensors in each BWTRG step.
The tensors away from the reflection line are updated in the same way as the bulk tensors in Appendix~\ref{appendix:bulk_tensor}.
Since the lattice is rotated by $45$ degrees in each step, the boundary tensors are converted alternately between $(\sigma_\Omega, A_{\Omega\sigma})$ and $(\tau_\Omega, A_{\Omega\tau})$.
Figure~\ref{fig:klein_update} illustrates the update from the former to the latter.
The reflection reverses the order of the bonds crossing the reflection line.
The reversal of $2^n$ bonds can be treated as swaps on the $n$ layers of a binary tree, i.e., the two halves are swapped and the bonds in each half are reversed recursively.
The crossing bonds in Fig.~\ref{fig:klein_update}(a) are drawn in this way: the two bonds in each pair are swapped, and the pairs are swapped with each other.
Since each pair of bonds is merged into a single bond in a BWTRG step, the swap in the lowest layer is absorbed into the boundary tensors, and only the remaining swaps act on the coarse-grained bonds as in Fig.~\ref{fig:klein_update}(b).

The plaquette on the reflection line in Fig.~\ref{fig:klein_update}(a) is decomposed in a similar manner to Eq.~\eqref{eq:plaquette_decomp} as
\begin{equation}
  \includetikz{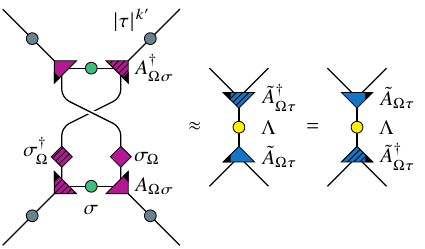}.
  \label{eq:klein_decomp}
\end{equation}
As in the bulk, the factors $|\tau|^{k'}$ are attached to the four external legs, while $\sigma_\Omega$ inside the plaquette is kept as it is.
The plaquette can be reshaped into a Hermitian matrix between the lower and upper pairs of the external legs, but its eigenvalues can be negative.
We retain up to the $\chi$ largest eigenvalues in absolute value and obtain the diagonal matrix $\Lambda$ of the eigenvalues and the isometry $\tilde{A}_{\Omega\tau}$ of the corresponding eigenvectors.

As shown in Fig.~\ref{fig:klein_update}(b), $\tau_\Omega$ is sandwiched between $A_{\Omega\tau}^\dagger$ on both sides, while $\tilde{A}_{\Omega\tau}$ and $\tilde{A}_{\Omega\tau}^\dagger$ appear on the two sides in Eq.~\eqref{eq:klein_decomp}.
We thus define $\tilde{\tau}_\Omega$ as
\begin{equation}
  \includetikz{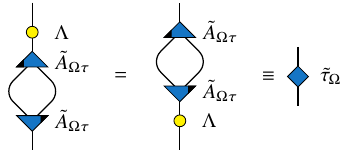},
  \label{eq:tau_omega}
\end{equation}
where $\Lambda$ can be placed on either of the two isometries owing to the last equality in Eq.~\eqref{eq:klein_decomp}.
Unlike the bond weights in the bulk, $\tilde{\tau}_\Omega$ is not diagonal in general.
Multiplying both sides of the last equality in Eq.~\eqref{eq:klein_decomp} by a projector $\tilde{A}_{\Omega\tau}\tilde{A}_{\Omega\tau}^\dagger$, we obtain
\begin{equation}
  \includetikz{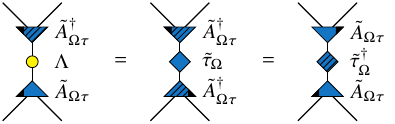},
  \label{eq:klein_decomp_2}
\end{equation}
where the same isometry appears on both sides of the reflection line.
The two expressions in Eq.~\eqref{eq:klein_decomp_2} appear alternately along the reflection line in Fig.~\ref{fig:klein_update}(b), and each is chosen to be consistent with the bulk tensors connected to it.

Next, we describe the update from $(\tau_\Omega, A_{\Omega\tau})$ to $(\sigma_\Omega, A_{\Omega\sigma})$.
If the plaquette containing $A_{\Omega\tau}$ were decomposed independently, the region of tensors different from the bulk ones would spread.
Instead, we update the boundary tensors by using $\tilde{A}$ and $\tilde{\tau}$ obtained in the bulk update so that we approximate the plaquette as
\begin{equation}
  \includetikz{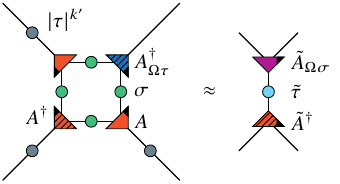},
  \label{eq:iso_omega_sigma}
\end{equation}
where the factors $|\tau|^{k'}$ are attached to the three external legs other than the one on the reflection line.
Multiplying both sides by $\tilde{A}$ and $\tilde{\tau}^{-1}$ from below and using the isometry condition, we obtain $\tilde{A}_{\Omega\sigma}$ as
\begin{equation}
  \includetikz{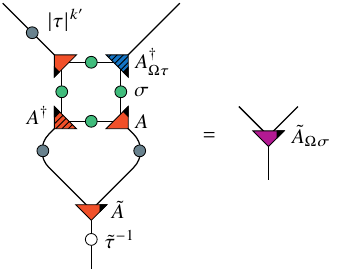}.
  \label{eq:iso_omega_sigma_2}
\end{equation}
Note that $\tilde{A}_{\Omega\sigma}$ is not an isometry in general, unlike $\tilde{A}_{\Omega\tau}$.
Since the factor $|\tau|^{k'}$ is not attached to the leg on the reflection line, $\tau_\Omega$ is used as it is in the next step:
\begin{equation}
  \tilde{\sigma}_\Omega = \tau_\Omega.
\end{equation}

The partition function on the Klein bottle is obtained by contracting the cluster of the same shape as Eq.~\eqref{eq:trace_16a}, in which the bonds in the middle are reconnected to the mirror-image positions through the boundary tensors:
\begin{equation}
  Z^{\mathcal{K},\text{TN}} = \includetikz{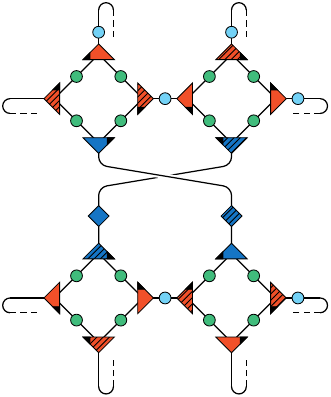},
  \label{eq:klein_16a}
\end{equation}
which corresponds to the Klein bottle of the same size $L\times L$ as the torus in Eq.~\eqref{eq:trace_16a}.
The reflection line through the $\tau$ bonds corresponds to a reflection about a diagonal at even steps and about a lattice axis at odd steps, from which we obtain $Z^{\mathcal{K}_D}$ and $Z^{\mathcal{K}_A}$, respectively.
As for the torus, $Z^{\mathcal{K}}_\text{pa}$ and $Z^{\mathcal{K}}_\text{pp}$ are extracted by fixing the $\mathbb{Z}_2$ parities of the bonds crossing the boundaries.

For the initial tensors of the present model, we set the initial boundary tensors as $A_{\Omega\sigma}=A_{\Omega\tau}=A$, and $\sigma_\Omega$ and $\tau_\Omega$ are given by $\sigma$ and $\tau$ with the flow of the conserved quantity through one of their indices reversed [see Eq.~\eqref{eq:A_flow}].
We keep both types of the boundary tensors and perform the two updates described above simultaneously in each step.

The additional computational cost for the boundary tensors is of the same order as that for the bulk tensors, since the decomposition in Eq.~\eqref{eq:klein_decomp} and the projection in Eq.~\eqref{eq:iso_omega_sigma_2} involve plaquettes of the same size as Eq.~\eqref{eq:plaquette_decomp}.
The contraction in Eq.~\eqref{eq:klein_16a} requires the $O(\chi^6)$ computational cost, the same as Eq.~\eqref{eq:trace_16a}.
In practice, the computation of the Klein-bottle partition functions roughly doubles the computational time per step.

\section{Partition functions of the \texorpdfstring{$\mathbb{Z}_2$}{Z2} orbifold CFT}
\label{appendix:cft}

In this appendix, we summarize the partition functions of the $\mathbb{Z}_2$ orbifold CFT on the torus and the Klein bottle, from which the universal values of the partition-function ratios in Sec.~\ref{sec:results} are obtained.
The critical line of the AT universality class is described by the $\mathbb{Z}_2$ orbifold, $\varphi\to -\varphi$, of the free boson compactified on a circle of radius $r$, $\varphi\equiv \varphi+2\pi r$~\cite{saleur1987partition,yang1987modular,ginsparg1988applied,ginsparg1988curiosities}.
We follow the convention of Refs.~\cite{ginsparg1988applied,ginsparg1988curiosities} for the radius, in which $r=1$ and $r=\sqrt{2}$ correspond to the $\text{Ising}^2$ and 4-state Potts CFTs, respectively.

The partition function of the compactified boson on a torus is given by
\begin{equation}
  \mathcal{Z}_\text{circ}(r) = \frac{1}{|\eta(\tau)|^2}\sum_{m,n\in\mathbb{Z}} q^{\frac12\left(\frac{m}{2r}+nr\right)^2}\,\bar{q}^{\frac12\left(\frac{m}{2r}-nr\right)^2},
\end{equation}
where $q=e^{2\pi i\tau}$ with the modular parameter $\tau$, and $\eta(\tau)=q^{1/24}\prod_{k\geq1}(1-q^k)$ is the Dedekind eta function.
The $\mathbb{Z}_2$ orbifold consists of the untwisted sector projected onto the states invariant under $\varphi\to -\varphi$ and the twisted sector with $\varphi(x+L)=-\varphi(x)$.
For the trace in Eq.~\eqref{eq:ZK_CFT}, the modular parameter is $\tau=i\ell/L$, i.e., $q=e^{-2\pi\ell/L}$.
We denote the partition function of the orbifold with $g$ inserted into the trace by $\mathcal{Z}[g]$, which is given by Eq.~\eqref{eq:ZK_CFT} without $\Omega$.
The modular-invariant partition function of the orbifold, $\mathcal{Z}[1]$, is
\begin{equation}
  \mathcal{Z}[1] = \frac{1}{2}\mathcal{Z}_\text{circ}(r)
  + \left|\frac{\eta}{\theta_2}\right|
  + \left|\frac{\eta}{\theta_3}\right|
  + \left|\frac{\eta}{\theta_4}\right|,
  \label{eq:Z_orb}
\end{equation}
with the Jacobi theta functions $\theta_2=\sum_{n}q^{(n+1/2)^2/2}$, $\theta_3=\sum_{n}q^{n^2/2}$, and $\theta_4=\sum_{n}(-1)^n q^{n^2/2}$.
The second term comes from the untwisted sector with $\varphi\to -\varphi$ inserted into the trace, and the last two terms from the twisted sector.
Since $d_1$ acts as $-1$ on the twisted sector of the orbifold, we obtain
\begin{equation}
  \mathcal{Z}[d_1] = \frac{1}{2}\mathcal{Z}_\text{circ}(r)
  + \left|\frac{\eta}{\theta_2}\right|
  - \left|\frac{\eta}{\theta_3}\right|
  - \left|\frac{\eta}{\theta_4}\right|.
  \label{eq:Z_d1_orb}
\end{equation}
The universal value of $R_\text{pa}$ in Sec.~\ref{sec:results_torus} is given by $\mathcal{Z}[d_1]/\mathcal{Z}[1]$ with $\ell=L$.

The Klein-bottle partition function $\mathcal{Z}[\Omega g]$ in Eq.~\eqref{eq:ZK_CFT} is also written as the sum of the contributions from the untwisted and twisted sectors~\cite{blumenhagen2009introduction,tu2017universal}.
For $g=1$, we have
\begin{equation}
  \mathcal{Z}[\Omega] = U + T,
\end{equation}
where
\begin{align}
  U &= \frac{1}{2\eta(2\tau)}\left[\sum_{m\in\mathbb{Z}} q^{m^2/4r^2} + \sum_{n\in\mathbb{Z}} q^{n^2r^2}\right], \\
  T &= \frac{2\,\eta(2\tau)}{\eta(\tau)}.
\end{align}
The first and second sums in $U$ come from the states with zero winding number and those with zero momentum, respectively, which are mapped to themselves by $\Omega$ in the orbifold.
As for the torus, $d_1$ reverses the sign of the contribution from the twisted sector:
\begin{equation}
  \mathcal{Z}[\Omega d_1] = U - T.
\end{equation}
On the other hand, $d_2$ acts on the untwisted sector as the half shift $\varphi\to \varphi+\pi r$, while it exchanges the two fixed points, $\varphi=0$ and $\pi r$, in the twisted sector.
Therefore, the contribution of the twisted sector vanishes, and we obtain
\begin{equation}
  \mathcal{Z}[\Omega d_2] = \frac{1}{2\eta(2\tau)}\left[\sum_{m\in\mathbb{Z}} (-1)^m q^{m^2/4r^2} + \sum_{n\in\mathbb{Z}} q^{n^2r^2}\right],
\end{equation}
where the phase $(-1)^m$ comes from the half shift.
At the Potts point $r=\sqrt{2}$, one can show $\mathcal{Z}[\Omega d_1] = \mathcal{Z}[\Omega d_2]$ by using the identity $\theta_2(\tau)=2\eta(2\tau)^2/\eta(\tau)$.
The universal values of $R^{\mathcal{K}_A}_\text{pa}$ and $R^{\mathcal{K}_D}_\text{pp}$ in Sec.~\ref{sec:results_klein} are given by $\mathcal{Z}[\Omega d_1]/\mathcal{Z}[1]$ and $\mathcal{Z}[\Omega d_2]/\mathcal{Z}[1]$ with $\ell=L$, respectively.

\begin{figure}[t]
  \centering
  \includegraphics{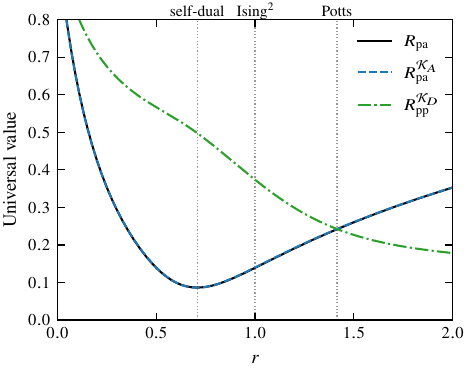}
  \caption{Universal values of the partition-function ratios $R_\text{pa}$, $R^{\mathcal{K}_A}_\text{pa}$, and $R^{\mathcal{K}_D}_\text{pp}$ of the $\mathbb{Z}_2$ orbifold CFT as functions of the compactification radius $r$, for the torus and the Klein bottles of size $L\times L$.
  The vertical dotted lines indicate the self-dual ($r=1/\sqrt{2}$), $\text{Ising}^2$ ($r=1$), and 4-state Potts ($r=\sqrt{2}$) points.
  The curves of $R_\text{pa}$ and $R^{\mathcal{K}_A}_\text{pa}$ almost overlap, since their difference is of the order of $10^{-3}$ or less.
  }
  \label{fig:cft_ratios}
\end{figure}
Figure~\ref{fig:cft_ratios} shows the universal values of the three ratios as functions of $r$.
The two ratios $R_\text{pa}$ and $R^{\mathcal{K}_A}_\text{pa}$ differ only by the contributions of the states exchanged by $\Omega$, which are smaller than the leading contributions by a factor of the order of $q=e^{-2\pi}\approx 2\times10^{-3}$ for $\ell=L$.
They are symmetric under the T-duality $r\leftrightarrow 1/(2r)$, which exchanges the momentum and winding states and leaves $\mathcal{Z}[1]$, $\mathcal{Z}[d_1]$, and $\mathcal{Z}[\Omega d_1]$ invariant.
In contrast, $R^{\mathcal{K}_D}_\text{pp}$ is not symmetric, since the half shift gives the phase $(-1)^m$ only to the momentum states.
$R^{\mathcal{K}_A}_\text{pa}$ and $R^{\mathcal{K}_D}_\text{pp}$ coincide only at the Potts point.
In the present model, the AT critical line for $g\geq g^*$ corresponds to $1<r\leq\sqrt{2}$, where $r=\sqrt{2}$ at $g=g^*$ and $r\to 1$ as $g\to\infty$.

\section{Transition temperatures}
\label{appendix:tc}

Table~\ref{tab:tc} lists the transition temperatures $T_\text{c}$ obtained by BWTRG with $\chi=192$.
The error is estimated from the difference between the results of $\chi=128$ and $192$.
The difference between $\chi=64$ and $128$ is one to two orders of magnitude larger, indicating the rapid convergence with respect to $\chi$.
Since $T_\text{c}$ converges with oscillation with respect to the bond dimension, the error may be underestimated.

\begin{table}[b]
  \centering
  \caption{Transition temperature $T_\text{c}$ obtained by BWTRG with $\chi=192$,
  compared with the Monte Carlo (MC)~\cite{li2024comprehensive} and transfer matrix (TM)~\cite{jin2013phase} results.
  }
  \label{tab:tc}
  \begin{ruledtabular}
  \begin{tabular}{llll}
  %\hline\hline
  $g$ & BWTRG ($\chi=192$) & MC~\cite{li2024comprehensive} & TM~\cite{jin2013phase} \\
  \hline
  0 & 2.26918532(2) & --- & --- \\
  0.2 & 1.60800041(1) & --- & --- \\
  0.4 & 0.867214600(1) & --- & --- \\
  0.55 & 0.771406(3) & --- & --- \\
  0.6 & 0.9716254(3) & 0.9717(2) & --- \\
  0.61 & 1.0066918(2) & --- & --- \\
  0.62 & 1.0407340(2) & 1.0408(3) & --- \\
  0.63 & 1.07389953(5) & --- & --- \\
  0.64 & 1.1063059(3) & 1.1063(2) & --- \\
  0.65 & 1.13804831(5) & --- & --- \\
  0.66 & 1.16920469(1) & --- & --- \\
  0.67 & 1.19984017(4) & --- & 1.1998(3) \\
  0.68 & 1.230009445(4) & --- & --- \\
  0.69 & 1.25975906(6) & --- & --- \\
  0.7 & 1.28912897(9) & 1.2892(2) & 1.2890(2) \\
  0.8 & 1.5677939(1) & --- & --- \\
  1 & 2.082099(1) & --- & 2.0821(2) \\
  %\hline\hline
  \end{tabular}
  \end{ruledtabular}
\end{table}

\newpage

\bibliography{j1j2_bwtrg}

%apsrev4-2.bst 2019-01-14 (MD) hand-edited version of apsrev4-1.bst
%Control: key (0)
%Control: author (8) initials jnrlst
%Control: editor formatted (1) identically to author
%Control: production of article title (0) allowed
%Control: page (0) single
%Control: year (1) truncated
%Control: production of eprint (1) enabled
\begin{thebibliography}{40}%
\makeatletter
\providecommand \@ifxundefined [1]{%
 \@ifx{#1\undefined}
}%
\providecommand \@ifnum [1]{%
 \ifnum #1\expandafter \@firstoftwo
 \else \expandafter \@secondoftwo
 \fi
}%
\providecommand \@ifx [1]{%
 \ifx #1\expandafter \@firstoftwo
 \else \expandafter \@secondoftwo
 \fi
}%
\providecommand \natexlab [1]{#1}%
\providecommand \enquote  [1]{``#1''}%
\providecommand \bibnamefont  [1]{#1}%
\providecommand \bibfnamefont [1]{#1}%
\providecommand \citenamefont [1]{#1}%
\providecommand \href@noop [0]{\@secondoftwo}%
\providecommand \href [0]{\begingroup \@sanitize@url \@href}%
\providecommand \@href[1]{\@@startlink{#1}\@@href}%
\providecommand \@@href[1]{\endgroup#1\@@endlink}%
\providecommand \@sanitize@url [0]{\catcode `\\12\catcode `\$12\catcode `\&12\catcode `\#12\catcode `\^12\catcode `\_12\catcode `\%12\relax}%
\providecommand \@@startlink[1]{}%
\providecommand \@@endlink[0]{}%
\providecommand \url  [0]{\begingroup\@sanitize@url \@url }%
\providecommand \@url [1]{\endgroup\@href {#1}{\urlprefix }}%
\providecommand \urlprefix  [0]{URL }%
\providecommand \Eprint [0]{\href }%
\providecommand \doibase [0]{https://doi.org/}%
\providecommand \selectlanguage [0]{\@gobble}%
\providecommand \bibinfo  [0]{\@secondoftwo}%
\providecommand \bibfield  [0]{\@secondoftwo}%
\providecommand \translation [1]{[#1]}%
\providecommand \BibitemOpen [0]{}%
\providecommand \bibitemStop [0]{}%
\providecommand \bibitemNoStop [0]{.\EOS\space}%
\providecommand \EOS [0]{\spacefactor3000\relax}%
\providecommand \BibitemShut  [1]{\csname bibitem#1\endcsname}%
\let\auto@bib@innerbib\@empty
%</preamble>
\bibitem [{\citenamefont {Diep}(2020)}]{diep2020frustrated}%
  \BibitemOpen
  \bibfield  {author} {\bibinfo {author} {\bibfnamefont {H.~T.}\ \bibnamefont {Diep}},\ }\href {https://doi.org/10.1142/11660} {\emph {\bibinfo {title} {Frustrated Spin Systems}}},\ \bibinfo {edition} {3rd}\ ed.\ (\bibinfo  {publisher} {World Scientific},\ \bibinfo {year} {2020})\BibitemShut {NoStop}%
\bibitem [{\citenamefont {Jin}\ \emph {et~al.}(2012)\citenamefont {Jin}, \citenamefont {Sen},\ and\ \citenamefont {Sandvik}}]{jin2012ashkinteller}%
  \BibitemOpen
  \bibfield  {author} {\bibinfo {author} {\bibfnamefont {S.}~\bibnamefont {Jin}}, \bibinfo {author} {\bibfnamefont {A.}~\bibnamefont {Sen}},\ and\ \bibinfo {author} {\bibfnamefont {A.~W.}\ \bibnamefont {Sandvik}},\ }\bibfield  {title} {\bibinfo {title} {Ashkin-{{Teller Criticality}} and {{Pseudo-First-Order Behavior}} in a {{Frustrated Ising Model}} on the {{Square Lattice}}},\ }\href {https://doi.org/10.1103/PhysRevLett.108.045702} {\bibfield  {journal} {\bibinfo  {journal} {Phys. Rev. Lett.}\ }\textbf {\bibinfo {volume} {108}},\ \bibinfo {pages} {045702} (\bibinfo {year} {2012})}\BibitemShut {NoStop}%
\bibitem [{\citenamefont {Kalz}\ and\ \citenamefont {Honecker}(2012)}]{kalz2012location}%
  \BibitemOpen
  \bibfield  {author} {\bibinfo {author} {\bibfnamefont {A.}~\bibnamefont {Kalz}}\ and\ \bibinfo {author} {\bibfnamefont {A.}~\bibnamefont {Honecker}},\ }\bibfield  {title} {\bibinfo {title} {Location of the {{Potts}}-critical end point in the frustrated {{Ising}} model on the square lattice},\ }\href {https://doi.org/10.1103/PhysRevB.86.134410} {\bibfield  {journal} {\bibinfo  {journal} {Phys. Rev. B}\ }\textbf {\bibinfo {volume} {86}},\ \bibinfo {pages} {134410} (\bibinfo {year} {2012})}\BibitemShut {NoStop}%
\bibitem [{\citenamefont {Jin}\ \emph {et~al.}(2013)\citenamefont {Jin}, \citenamefont {Sen}, \citenamefont {Guo},\ and\ \citenamefont {Sandvik}}]{jin2013phase}%
  \BibitemOpen
  \bibfield  {author} {\bibinfo {author} {\bibfnamefont {S.}~\bibnamefont {Jin}}, \bibinfo {author} {\bibfnamefont {A.}~\bibnamefont {Sen}}, \bibinfo {author} {\bibfnamefont {W.}~\bibnamefont {Guo}},\ and\ \bibinfo {author} {\bibfnamefont {A.~W.}\ \bibnamefont {Sandvik}},\ }\bibfield  {title} {\bibinfo {title} {Phase transitions in the frustrated {{Ising}} model on the square lattice},\ }\href {https://doi.org/10.1103/PhysRevB.87.144406} {\bibfield  {journal} {\bibinfo  {journal} {Phys. Rev. B}\ }\textbf {\bibinfo {volume} {87}},\ \bibinfo {pages} {144406} (\bibinfo {year} {2013})}\BibitemShut {NoStop}%
\bibitem [{\citenamefont {Ashkin}\ and\ \citenamefont {Teller}(1943)}]{ashkin1943statistics}%
  \BibitemOpen
  \bibfield  {author} {\bibinfo {author} {\bibfnamefont {J.}~\bibnamefont {Ashkin}}\ and\ \bibinfo {author} {\bibfnamefont {E.}~\bibnamefont {Teller}},\ }\bibfield  {title} {\bibinfo {title} {Statistics of {{Two-Dimensional Lattices}} with {{Four Components}}},\ }\href {https://doi.org/10.1103/PhysRev.64.178} {\bibfield  {journal} {\bibinfo  {journal} {Phys. Rev.}\ }\textbf {\bibinfo {volume} {64}},\ \bibinfo {pages} {178} (\bibinfo {year} {1943})}\BibitemShut {NoStop}%
\bibitem [{\citenamefont {Delfino}\ and\ \citenamefont {Grinza}(2004)}]{delfino2004universal}%
  \BibitemOpen
  \bibfield  {author} {\bibinfo {author} {\bibfnamefont {G.}~\bibnamefont {Delfino}}\ and\ \bibinfo {author} {\bibfnamefont {P.}~\bibnamefont {Grinza}},\ }\bibfield  {title} {\bibinfo {title} {Universal ratios along a line of critical points. {{The Ashkin}}--{{Teller}} model},\ }\href {https://doi.org/10.1016/j.nuclphysb.2004.01.007} {\bibfield  {journal} {\bibinfo  {journal} {Nucl. Phys. B}\ }\textbf {\bibinfo {volume} {682}},\ \bibinfo {pages} {521} (\bibinfo {year} {2004})}\BibitemShut {NoStop}%
\bibitem [{\citenamefont {Li}\ and\ \citenamefont {Jiang}(2024)}]{li2024comprehensive}%
  \BibitemOpen
  \bibfield  {author} {\bibinfo {author} {\bibfnamefont {S.-W.}\ \bibnamefont {Li}}\ and\ \bibinfo {author} {\bibfnamefont {F.-J.}\ \bibnamefont {Jiang}},\ }\bibfield  {title} {\bibinfo {title} {A comprehensive study of the phase transitions of the frustrated {$J_1$}-{$J_2$} {{Ising}} model on the square lattice},\ }\href {https://doi.org/10.1093/ptep/ptae061} {\bibfield  {journal} {\bibinfo  {journal} {Prog. Theor. Exp. Phys.}\ }\textbf {\bibinfo {volume} {2024}},\ \bibinfo {pages} {053A06} (\bibinfo {year} {2024})}\BibitemShut {NoStop}%
\bibitem [{\citenamefont {Sun}\ and\ \citenamefont {Fan}(2026)}]{sun2026relevantmarginal}%
  \BibitemOpen
  \bibfield  {author} {\bibinfo {author} {\bibfnamefont {Y.}~\bibnamefont {Sun}}\ and\ \bibinfo {author} {\bibfnamefont {Y.}~\bibnamefont {Fan}},\ }\bibfield  {title} {\bibinfo {title} {Relevant-marginal field mixing resolves the four-state {{Potts}} endpoint in the square-lattice {${J}_{1}\text{\ensuremath{-}}{J}_{2}$} {{Ising}} model},\ }\href {https://doi.org/10.1103/8yb7-rs6t} {\bibfield  {journal} {\bibinfo  {journal} {Phys. Rev. B}\ }\textbf {\bibinfo {volume} {114}},\ \bibinfo {pages} {L080406} (\bibinfo {year} {2026})}\BibitemShut {NoStop}%
\bibitem [{\citenamefont {Levin}\ and\ \citenamefont {Nave}(2007)}]{levin2007tensor}%
  \BibitemOpen
  \bibfield  {author} {\bibinfo {author} {\bibfnamefont {M.}~\bibnamefont {Levin}}\ and\ \bibinfo {author} {\bibfnamefont {C.~P.}\ \bibnamefont {Nave}},\ }\bibfield  {title} {\bibinfo {title} {Tensor {{Renormalization Group Approach}} to {{Two-Dimensional Classical Lattice Models}}},\ }\href {https://doi.org/10.1103/PhysRevLett.99.120601} {\bibfield  {journal} {\bibinfo  {journal} {Phys. Rev. Lett.}\ }\textbf {\bibinfo {volume} {99}},\ \bibinfo {pages} {120601} (\bibinfo {year} {2007})}\BibitemShut {NoStop}%
\bibitem [{\citenamefont {Xie}\ \emph {et~al.}(2012)\citenamefont {Xie}, \citenamefont {Chen}, \citenamefont {Qin}, \citenamefont {Zhu}, \citenamefont {Yang},\ and\ \citenamefont {Xiang}}]{xie2012coarsegraining}%
  \BibitemOpen
  \bibfield  {author} {\bibinfo {author} {\bibfnamefont {Z.~Y.}\ \bibnamefont {Xie}}, \bibinfo {author} {\bibfnamefont {J.}~\bibnamefont {Chen}}, \bibinfo {author} {\bibfnamefont {M.~P.}\ \bibnamefont {Qin}}, \bibinfo {author} {\bibfnamefont {J.~W.}\ \bibnamefont {Zhu}}, \bibinfo {author} {\bibfnamefont {L.~P.}\ \bibnamefont {Yang}},\ and\ \bibinfo {author} {\bibfnamefont {T.}~\bibnamefont {Xiang}},\ }\bibfield  {title} {\bibinfo {title} {Coarse-graining renormalization by higher-order singular value decomposition},\ }\href {https://doi.org/10.1103/PhysRevB.86.045139} {\bibfield  {journal} {\bibinfo  {journal} {Phys. Rev. B}\ }\textbf {\bibinfo {volume} {86}},\ \bibinfo {pages} {045139} (\bibinfo {year} {2012})}\BibitemShut {NoStop}%
\bibitem [{\citenamefont {Gangat}\ and\ \citenamefont {Kao}(2019)}]{gangat2019phase}%
  \BibitemOpen
  \bibfield  {author} {\bibinfo {author} {\bibfnamefont {A.~A.}\ \bibnamefont {Gangat}}\ and\ \bibinfo {author} {\bibfnamefont {Y.-J.}\ \bibnamefont {Kao}},\ }\bibfield  {title} {\bibinfo {title} {Phase boundary location with information-theoretic entropy in tensor renormalization group flows},\ }\href {https://doi.org/10.1103/PhysRevB.100.094430} {\bibfield  {journal} {\bibinfo  {journal} {Phys. Rev. B}\ }\textbf {\bibinfo {volume} {100}},\ \bibinfo {pages} {094430} (\bibinfo {year} {2019})}\BibitemShut {NoStop}%
\bibitem [{\citenamefont {Yoshiyama}\ and\ \citenamefont {Hukushima}(2023)}]{yoshiyama2023higherordera}%
  \BibitemOpen
  \bibfield  {author} {\bibinfo {author} {\bibfnamefont {K.}~\bibnamefont {Yoshiyama}}\ and\ \bibinfo {author} {\bibfnamefont {K.}~\bibnamefont {Hukushima}},\ }\bibfield  {title} {\bibinfo {title} {Higher-order tensor renormalization group study of the {${J}_{1}\text{\ensuremath{-}}{J}_{2}$} {{Ising}} model on a square lattice},\ }\href {https://doi.org/10.1103/PhysRevE.108.054124} {\bibfield  {journal} {\bibinfo  {journal} {Phys. Rev. E}\ }\textbf {\bibinfo {volume} {108}},\ \bibinfo {pages} {054124} (\bibinfo {year} {2023})}\BibitemShut {NoStop}%
\bibitem [{\citenamefont {Li}\ and\ \citenamefont {Yang}(2021)}]{li2021tensor}%
  \BibitemOpen
  \bibfield  {author} {\bibinfo {author} {\bibfnamefont {H.}~\bibnamefont {Li}}\ and\ \bibinfo {author} {\bibfnamefont {L.-P.}\ \bibnamefont {Yang}},\ }\bibfield  {title} {\bibinfo {title} {Tensor network simulation for the frustrated {${J}_{1}\text{\ensuremath{-}}{J}_{2}$} {{Ising}} model on the square lattice},\ }\href {https://doi.org/10.1103/PhysRevE.104.024118} {\bibfield  {journal} {\bibinfo  {journal} {Phys. Rev. E}\ }\textbf {\bibinfo {volume} {104}},\ \bibinfo {pages} {024118} (\bibinfo {year} {2021})}\BibitemShut {NoStop}%
\bibitem [{\citenamefont {Gangat}(2024)}]{gangat2024weak}%
  \BibitemOpen
  \bibfield  {author} {\bibinfo {author} {\bibfnamefont {A.~A.}\ \bibnamefont {Gangat}},\ }\bibfield  {title} {\bibinfo {title} {Weak first-order phase transitions in the frustrated square lattice {${J}_{1}\text{\ensuremath{-}}{J}_{2}$} classical {{Ising}} model},\ }\href {https://doi.org/10.1103/PhysRevB.109.104419} {\bibfield  {journal} {\bibinfo  {journal} {Phys. Rev. B}\ }\textbf {\bibinfo {volume} {109}},\ \bibinfo {pages} {104419} (\bibinfo {year} {2024})}\BibitemShut {NoStop}%
\bibitem [{\citenamefont {Adachi}\ \emph {et~al.}(2022)\citenamefont {Adachi}, \citenamefont {Okubo},\ and\ \citenamefont {Todo}}]{adachi2022bondweighted}%
  \BibitemOpen
  \bibfield  {author} {\bibinfo {author} {\bibfnamefont {D.}~\bibnamefont {Adachi}}, \bibinfo {author} {\bibfnamefont {T.}~\bibnamefont {Okubo}},\ and\ \bibinfo {author} {\bibfnamefont {S.}~\bibnamefont {Todo}},\ }\bibfield  {title} {\bibinfo {title} {Bond-weighted tensor renormalization group},\ }\href {https://doi.org/10.1103/PhysRevB.105.L060402} {\bibfield  {journal} {\bibinfo  {journal} {Phys. Rev. B}\ }\textbf {\bibinfo {volume} {105}},\ \bibinfo {pages} {L060402} (\bibinfo {year} {2022})}\BibitemShut {NoStop}%
\bibitem [{\citenamefont {Morita}\ and\ \citenamefont {Kawashima}(2025)}]{morita2025multiimpurity}%
  \BibitemOpen
  \bibfield  {author} {\bibinfo {author} {\bibfnamefont {S.}~\bibnamefont {Morita}}\ and\ \bibinfo {author} {\bibfnamefont {N.}~\bibnamefont {Kawashima}},\ }\bibfield  {title} {\bibinfo {title} {Multi-impurity method for the bond-weighted tensor renormalization group},\ }\href {https://doi.org/10.1103/PhysRevB.111.054433} {\bibfield  {journal} {\bibinfo  {journal} {Phys. Rev. B}\ }\textbf {\bibinfo {volume} {111}},\ \bibinfo {pages} {054433} (\bibinfo {year} {2025})}\BibitemShut {NoStop}%
\bibitem [{\citenamefont {Lyu}(2026)}]{lyu2026latticea}%
  \BibitemOpen
  \bibfield  {author} {\bibinfo {author} {\bibfnamefont {X.}~\bibnamefont {Lyu}},\ }\bibfield  {title} {\bibinfo {title} {Lattice and {$\mathcal{PT}$} symmetries in tensor-network renormalization group: {{Case}} study of a hard-square lattice gas model},\ }\href {https://doi.org/10.1103/vl23-xsyr} {\bibfield  {journal} {\bibinfo  {journal} {Phys. Rev. B}\ }\textbf {\bibinfo {volume} {114}},\ \bibinfo {pages} {034111} (\bibinfo {year} {2026})}\BibitemShut {NoStop}%
\bibitem [{\citenamefont {Ginsparg}(1990)}]{ginsparg1988applied}%
  \BibitemOpen
  \bibfield  {author} {\bibinfo {author} {\bibfnamefont {P.}~\bibnamefont {Ginsparg}},\ }\bibfield  {title} {\bibinfo {title} {Applied conformal field theory},\ }in\ \href@noop {} {\emph {\bibinfo {booktitle} {Fields, Strings and Critical Phenomena (Les Houches 1988)}}},\ \bibinfo {editor} {edited by\ \bibinfo {editor} {\bibfnamefont {E.}~\bibnamefont {Br{\'e}zin}}\ and\ \bibinfo {editor} {\bibfnamefont {J.}~\bibnamefont {{Zinn-Justin}}}}\ (\bibinfo  {publisher} {North-Holland},\ \bibinfo {address} {Amsterdam},\ \bibinfo {year} {1990})\ pp.\ \bibinfo {pages} {1--168},\ \Eprint {https://arxiv.org/abs/hep-th/9108028} {hep-th/9108028} \BibitemShut {NoStop}%
\bibitem [{\citenamefont {Di~Francesco}\ \emph {et~al.}(1997)\citenamefont {Di~Francesco}, \citenamefont {Mathieu},\ and\ \citenamefont {S{\'e}n{\'e}chal}}]{difrancesco1997conformal}%
  \BibitemOpen
  \bibfield  {author} {\bibinfo {author} {\bibfnamefont {P.}~\bibnamefont {Di~Francesco}}, \bibinfo {author} {\bibfnamefont {P.}~\bibnamefont {Mathieu}},\ and\ \bibinfo {author} {\bibfnamefont {D.}~\bibnamefont {S{\'e}n{\'e}chal}},\ }\href {https://doi.org/10.1007/978-1-4612-2256-9} {\emph {\bibinfo {title} {Conformal {{Field Theory}}}}},\ Graduate {{Texts}} in {{Contemporary Physics}}\ (\bibinfo  {publisher} {Springer},\ \bibinfo {address} {New York, NY},\ \bibinfo {year} {1997})\BibitemShut {NoStop}%
\bibitem [{\citenamefont {Henkel}(1999)}]{henkel1999conformal}%
  \BibitemOpen
  \bibfield  {author} {\bibinfo {author} {\bibfnamefont {M.}~\bibnamefont {Henkel}},\ }\href {https://doi.org/10.1007/978-3-662-03937-3} {\emph {\bibinfo {title} {Conformal {{Invariance}} and {{Critical Phenomena}}}}}\ (\bibinfo  {publisher} {Springer},\ \bibinfo {address} {Berlin, Heidelberg},\ \bibinfo {year} {1999})\BibitemShut {NoStop}%
\bibitem [{\citenamefont {Saleur}(1987)}]{saleur1987partition}%
  \BibitemOpen
  \bibfield  {author} {\bibinfo {author} {\bibfnamefont {H.}~\bibnamefont {Saleur}},\ }\bibfield  {title} {\bibinfo {title} {Partition functions of the two-dimensional {{Ashkin-Teller}} model on the critical line},\ }\href {https://doi.org/10.1088/0305-4470/20/16/016} {\bibfield  {journal} {\bibinfo  {journal} {J. Phys. A: Math. Gen.}\ }\textbf {\bibinfo {volume} {20}},\ \bibinfo {pages} {L1127} (\bibinfo {year} {1987})}\BibitemShut {NoStop}%
\bibitem [{\citenamefont {Ginsparg}(1988)}]{ginsparg1988curiosities}%
  \BibitemOpen
  \bibfield  {author} {\bibinfo {author} {\bibfnamefont {P.}~\bibnamefont {Ginsparg}},\ }\bibfield  {title} {\bibinfo {title} {Curiosities at {\emph{c}} = 1},\ }\href {https://doi.org/10.1016/0550-3213(88)90249-0} {\bibfield  {journal} {\bibinfo  {journal} {Nucl. Phys. B}\ }\textbf {\bibinfo {volume} {295}},\ \bibinfo {pages} {153} (\bibinfo {year} {1988})}\BibitemShut {NoStop}%
\bibitem [{\citenamefont {Tu}(2017)}]{tu2017universal}%
  \BibitemOpen
  \bibfield  {author} {\bibinfo {author} {\bibfnamefont {H.-H.}\ \bibnamefont {Tu}},\ }\bibfield  {title} {\bibinfo {title} {Universal entropy of conformal critical theories on a klein bottle},\ }\href {https://doi.org/10.1103/PhysRevLett.119.261603} {\bibfield  {journal} {\bibinfo  {journal} {Phys. Rev. Lett.}\ }\textbf {\bibinfo {volume} {119}},\ \bibinfo {pages} {261603} (\bibinfo {year} {2017})}\BibitemShut {NoStop}%
\bibitem [{\citenamefont {Tang}\ \emph {et~al.}(2017)\citenamefont {Tang}, \citenamefont {Chen}, \citenamefont {Li}, \citenamefont {Xie}, \citenamefont {Tu},\ and\ \citenamefont {Wang}}]{tang2017universal}%
  \BibitemOpen
  \bibfield  {author} {\bibinfo {author} {\bibfnamefont {W.}~\bibnamefont {Tang}}, \bibinfo {author} {\bibfnamefont {L.}~\bibnamefont {Chen}}, \bibinfo {author} {\bibfnamefont {W.}~\bibnamefont {Li}}, \bibinfo {author} {\bibfnamefont {X.~C.}\ \bibnamefont {Xie}}, \bibinfo {author} {\bibfnamefont {H.-H.}\ \bibnamefont {Tu}},\ and\ \bibinfo {author} {\bibfnamefont {L.}~\bibnamefont {Wang}},\ }\bibfield  {title} {\bibinfo {title} {Universal boundary entropies in conformal field theory: {{A}} quantum {{Monte Carlo}} study},\ }\href {https://doi.org/10.1103/PhysRevB.96.115136} {\bibfield  {journal} {\bibinfo  {journal} {Phys. Rev. B}\ }\textbf {\bibinfo {volume} {96}},\ \bibinfo {pages} {115136} (\bibinfo {year} {2017})}\BibitemShut {NoStop}%
\bibitem [{\citenamefont {Chen}\ \emph {et~al.}(2017)\citenamefont {Chen}, \citenamefont {Wang}, \citenamefont {Wang},\ and\ \citenamefont {Li}}]{chen2017conformal}%
  \BibitemOpen
  \bibfield  {author} {\bibinfo {author} {\bibfnamefont {L.}~\bibnamefont {Chen}}, \bibinfo {author} {\bibfnamefont {H.-X.}\ \bibnamefont {Wang}}, \bibinfo {author} {\bibfnamefont {L.}~\bibnamefont {Wang}},\ and\ \bibinfo {author} {\bibfnamefont {W.}~\bibnamefont {Li}},\ }\bibfield  {title} {\bibinfo {title} {Conformal thermal tensor network and universal entropy on topological manifolds},\ }\href {https://doi.org/10.1103/PhysRevB.96.174429} {\bibfield  {journal} {\bibinfo  {journal} {Phys. Rev. B}\ }\textbf {\bibinfo {volume} {96}},\ \bibinfo {pages} {174429} (\bibinfo {year} {2017})}\BibitemShut {NoStop}%
\bibitem [{\citenamefont {Tang}\ \emph {et~al.}(2019)\citenamefont {Tang}, \citenamefont {Xie}, \citenamefont {Wang},\ and\ \citenamefont {Tu}}]{tang2019klein}%
  \BibitemOpen
  \bibfield  {author} {\bibinfo {author} {\bibfnamefont {W.}~\bibnamefont {Tang}}, \bibinfo {author} {\bibfnamefont {X.~C.}\ \bibnamefont {Xie}}, \bibinfo {author} {\bibfnamefont {L.}~\bibnamefont {Wang}},\ and\ \bibinfo {author} {\bibfnamefont {H.-H.}\ \bibnamefont {Tu}},\ }\bibfield  {title} {\bibinfo {title} {Klein bottle entropy of compactified boson conformal field theory},\ }\href {https://doi.org/10.1103/PhysRevB.99.115105} {\bibfield  {journal} {\bibinfo  {journal} {Phys. Rev. B}\ }\textbf {\bibinfo {volume} {99}},\ \bibinfo {pages} {115105} (\bibinfo {year} {2019})}\BibitemShut {NoStop}%
\bibitem [{\citenamefont {Zhang}\ \emph {et~al.}(2023)\citenamefont {Zhang}, \citenamefont {Hulsch}, \citenamefont {Zhang}, \citenamefont {Tang}, \citenamefont {Wang},\ and\ \citenamefont {Tu}}]{zhang2023universal}%
  \BibitemOpen
  \bibfield  {author} {\bibinfo {author} {\bibfnamefont {Y.}~\bibnamefont {Zhang}}, \bibinfo {author} {\bibfnamefont {A.}~\bibnamefont {Hulsch}}, \bibinfo {author} {\bibfnamefont {H.-C.}\ \bibnamefont {Zhang}}, \bibinfo {author} {\bibfnamefont {W.}~\bibnamefont {Tang}}, \bibinfo {author} {\bibfnamefont {L.}~\bibnamefont {Wang}},\ and\ \bibinfo {author} {\bibfnamefont {H.-H.}\ \bibnamefont {Tu}},\ }\bibfield  {title} {\bibinfo {title} {Universal {{Scaling}} of {{Klein Bottle Entropy}} near {{Conformal Critical Points}}},\ }\href {https://doi.org/10.1103/PhysRevLett.130.151602} {\bibfield  {journal} {\bibinfo  {journal} {Phys. Rev. Lett.}\ }\textbf {\bibinfo {volume} {130}},\ \bibinfo {pages} {151602} (\bibinfo {year} {2023})}\BibitemShut {NoStop}%
\bibitem [{\citenamefont {Wang}\ and\ \citenamefont {Yang}(2026)}]{wang2026kleina}%
  \BibitemOpen
  \bibfield  {author} {\bibinfo {author} {\bibfnamefont {Z.-H.}\ \bibnamefont {Wang}}\ and\ \bibinfo {author} {\bibfnamefont {L.-P.}\ \bibnamefont {Yang}},\ }\bibfield  {title} {\bibinfo {title} {Klein bottle ratio of two-dimensional ferromagnetic {{Potts}} models},\ }\href {https://doi.org/10.1103/rnrc-m25p} {\bibfield  {journal} {\bibinfo  {journal} {Phys. Rev. E}\ }\textbf {\bibinfo {volume} {114}},\ \bibinfo {pages} {054115} (\bibinfo {year} {2026})}\BibitemShut {NoStop}%
\bibitem [{\citenamefont {Shimizu}\ and\ \citenamefont {Ueda}(2026)}]{shimizu2026tensor}%
  \BibitemOpen
  \bibfield  {author} {\bibinfo {author} {\bibfnamefont {H.}~\bibnamefont {Shimizu}}\ and\ \bibinfo {author} {\bibfnamefont {A.}~\bibnamefont {Ueda}},\ }\bibfield  {title} {\bibinfo {title} {Tensor network simulations for nonorientable surfaces},\ }\href {https://doi.org/10.1103/t4j5-qc4y} {\bibfield  {journal} {\bibinfo  {journal} {Phys. Rev. E}\ }\textbf {\bibinfo {volume} {113}},\ \bibinfo {pages} {034109} (\bibinfo {year} {2026})}\BibitemShut {NoStop}%
\bibitem [{\citenamefont {Zhao}\ \emph {et~al.}(2010)\citenamefont {Zhao}, \citenamefont {Xie}, \citenamefont {Chen}, \citenamefont {Wei}, \citenamefont {Cai},\ and\ \citenamefont {Xiang}}]{zhao2010renormalization}%
  \BibitemOpen
  \bibfield  {author} {\bibinfo {author} {\bibfnamefont {H.~H.}\ \bibnamefont {Zhao}}, \bibinfo {author} {\bibfnamefont {Z.~Y.}\ \bibnamefont {Xie}}, \bibinfo {author} {\bibfnamefont {Q.~N.}\ \bibnamefont {Chen}}, \bibinfo {author} {\bibfnamefont {Z.~C.}\ \bibnamefont {Wei}}, \bibinfo {author} {\bibfnamefont {J.~W.}\ \bibnamefont {Cai}},\ and\ \bibinfo {author} {\bibfnamefont {T.}~\bibnamefont {Xiang}},\ }\bibfield  {title} {\bibinfo {title} {Renormalization of tensor-network states},\ }\href {https://doi.org/10.1103/PhysRevB.81.174411} {\bibfield  {journal} {\bibinfo  {journal} {Phys. Rev. B}\ }\textbf {\bibinfo {volume} {81}},\ \bibinfo {pages} {174411} (\bibinfo {year} {2010})}\BibitemShut {NoStop}%
\bibitem [{\citenamefont {Singh}\ \emph {et~al.}(2010)\citenamefont {Singh}, \citenamefont {Pfeifer},\ and\ \citenamefont {Vidal}}]{singh2010tensor}%
  \BibitemOpen
  \bibfield  {author} {\bibinfo {author} {\bibfnamefont {S.}~\bibnamefont {Singh}}, \bibinfo {author} {\bibfnamefont {R.~N.~C.}\ \bibnamefont {Pfeifer}},\ and\ \bibinfo {author} {\bibfnamefont {G.}~\bibnamefont {Vidal}},\ }\bibfield  {title} {\bibinfo {title} {Tensor network decompositions in the presence of a global symmetry},\ }\href {https://doi.org/10.1103/PhysRevA.82.050301} {\bibfield  {journal} {\bibinfo  {journal} {Phys. Rev. A}\ }\textbf {\bibinfo {volume} {82}},\ \bibinfo {pages} {050301(R)} (\bibinfo {year} {2010})}\BibitemShut {NoStop}%
\bibitem [{\citenamefont {Singh}\ \emph {et~al.}(2011)\citenamefont {Singh}, \citenamefont {Pfeifer},\ and\ \citenamefont {Vidal}}]{singh2011tensor}%
  \BibitemOpen
  \bibfield  {author} {\bibinfo {author} {\bibfnamefont {S.}~\bibnamefont {Singh}}, \bibinfo {author} {\bibfnamefont {R.~N.~C.}\ \bibnamefont {Pfeifer}},\ and\ \bibinfo {author} {\bibfnamefont {G.}~\bibnamefont {Vidal}},\ }\bibfield  {title} {\bibinfo {title} {Tensor network states and algorithms in the presence of a global {{U}}(1) symmetry},\ }\href {https://doi.org/10.1103/PhysRevB.83.115125} {\bibfield  {journal} {\bibinfo  {journal} {Phys. Rev. B}\ }\textbf {\bibinfo {volume} {83}},\ \bibinfo {pages} {115125} (\bibinfo {year} {2011})}\BibitemShut {NoStop}%
\bibitem [{\citenamefont {Gu}\ and\ \citenamefont {Wen}(2009)}]{gu2009tensorentanglementfiltering}%
  \BibitemOpen
  \bibfield  {author} {\bibinfo {author} {\bibfnamefont {Z.~C.}\ \bibnamefont {Gu}}\ and\ \bibinfo {author} {\bibfnamefont {X.~G.}\ \bibnamefont {Wen}},\ }\bibfield  {title} {\bibinfo {title} {Tensor-entanglement-filtering renormalization approach and symmetry-protected topological order},\ }\href {https://doi.org/10.1103/PhysRevB.80.155131} {\bibfield  {journal} {\bibinfo  {journal} {Phys. Rev. B}\ }\textbf {\bibinfo {volume} {80}},\ \bibinfo {pages} {155131} (\bibinfo {year} {2009})}\BibitemShut {NoStop}%
\bibitem [{\citenamefont {Morita}\ and\ \citenamefont {Kawashima}(2026)}]{morita2026tensor}%
  \BibitemOpen
  \bibfield  {author} {\bibinfo {author} {\bibfnamefont {S.}~\bibnamefont {Morita}}\ and\ \bibinfo {author} {\bibfnamefont {N.}~\bibnamefont {Kawashima}},\ }\bibfield  {title} {\bibinfo {title} {Tensor {{Renormalization Group Calculations}} of {{Partition-Function Ratios}}},\ }\href {https://doi.org/10.7566/JPSJ.95.044001} {\bibfield  {journal} {\bibinfo  {journal} {J. Phys. Soc. Jpn.}\ }\textbf {\bibinfo {volume} {95}},\ \bibinfo {pages} {044001} (\bibinfo {year} {2026})}\BibitemShut {NoStop}%
\bibitem [{\citenamefont {Evenbly}\ and\ \citenamefont {Vidal}(2015)}]{evenbly2015tensor}%
  \BibitemOpen
  \bibfield  {author} {\bibinfo {author} {\bibfnamefont {G.}~\bibnamefont {Evenbly}}\ and\ \bibinfo {author} {\bibfnamefont {G.}~\bibnamefont {Vidal}},\ }\bibfield  {title} {\bibinfo {title} {Tensor {{Network Renormalization}}},\ }\href {https://doi.org/10.1103/PhysRevLett.115.180405} {\bibfield  {journal} {\bibinfo  {journal} {Phys. Rev. Lett.}\ }\textbf {\bibinfo {volume} {115}},\ \bibinfo {pages} {180405} (\bibinfo {year} {2015})}\BibitemShut {NoStop}%
\bibitem [{\citenamefont {Yang}\ \emph {et~al.}(2017)\citenamefont {Yang}, \citenamefont {Gu},\ and\ \citenamefont {Wen}}]{yang2017loop}%
  \BibitemOpen
  \bibfield  {author} {\bibinfo {author} {\bibfnamefont {S.}~\bibnamefont {Yang}}, \bibinfo {author} {\bibfnamefont {Z.-C.}\ \bibnamefont {Gu}},\ and\ \bibinfo {author} {\bibfnamefont {X.-G.}\ \bibnamefont {Wen}},\ }\bibfield  {title} {\bibinfo {title} {Loop {{Optimization}} for {{Tensor Network Renormalization}}},\ }\href {https://doi.org/10.1103/PhysRevLett.118.110504} {\bibfield  {journal} {\bibinfo  {journal} {Phys. Rev. Lett.}\ }\textbf {\bibinfo {volume} {118}},\ \bibinfo {pages} {110504} (\bibinfo {year} {2017})}\BibitemShut {NoStop}%
\bibitem [{\citenamefont {Hauru}\ \emph {et~al.}(2018)\citenamefont {Hauru}, \citenamefont {Delcamp},\ and\ \citenamefont {Mizera}}]{hauru2018renormalization}%
  \BibitemOpen
  \bibfield  {author} {\bibinfo {author} {\bibfnamefont {M.}~\bibnamefont {Hauru}}, \bibinfo {author} {\bibfnamefont {C.}~\bibnamefont {Delcamp}},\ and\ \bibinfo {author} {\bibfnamefont {S.}~\bibnamefont {Mizera}},\ }\bibfield  {title} {\bibinfo {title} {Renormalization of tensor networks using graph-independent local truncations},\ }\href {https://doi.org/10.1103/PhysRevB.97.045111} {\bibfield  {journal} {\bibinfo  {journal} {Phys. Rev. B}\ }\textbf {\bibinfo {volume} {97}},\ \bibinfo {pages} {045111} (\bibinfo {year} {2018})}\BibitemShut {NoStop}%
\bibitem [{\citenamefont {Harada}(2018)}]{harada2018entanglement}%
  \BibitemOpen
  \bibfield  {author} {\bibinfo {author} {\bibfnamefont {K.}~\bibnamefont {Harada}},\ }\bibfield  {title} {\bibinfo {title} {Entanglement branching operator},\ }\href {https://doi.org/10.1103/PhysRevB.97.045124} {\bibfield  {journal} {\bibinfo  {journal} {Phys. Rev. B}\ }\textbf {\bibinfo {volume} {97}},\ \bibinfo {pages} {045124} (\bibinfo {year} {2018})}\BibitemShut {NoStop}%
\bibitem [{\citenamefont {Yang}(1987)}]{yang1987modular}%
  \BibitemOpen
  \bibfield  {author} {\bibinfo {author} {\bibfnamefont {S.-K.}\ \bibnamefont {Yang}},\ }\bibfield  {title} {\bibinfo {title} {Modular invariant partition function of the {{Ashkin-Teller}} model on the critical line and {$N=2$} superconformal invariance},\ }\href {https://doi.org/10.1016/0550-3213(87)90334-8} {\bibfield  {journal} {\bibinfo  {journal} {Nucl. Phys. B}\ }\textbf {\bibinfo {volume} {285}},\ \bibinfo {pages} {183} (\bibinfo {year} {1987})}\BibitemShut {NoStop}%
\bibitem [{\citenamefont {Blumenhagen}\ and\ \citenamefont {Plauschinn}(2009)}]{blumenhagen2009introduction}%
  \BibitemOpen
  \bibfield  {author} {\bibinfo {author} {\bibfnamefont {R.}~\bibnamefont {Blumenhagen}}\ and\ \bibinfo {author} {\bibfnamefont {E.}~\bibnamefont {Plauschinn}},\ }\href {https://doi.org/10.1007/978-3-642-00450-6} {\emph {\bibinfo {title} {Introduction to {{Conformal Field Theory}}: {{With Applications}} to {{String Theory}}}}},\ \bibinfo {series} {Lecture {{Notes}} in {{Physics}}}, Vol.\ \bibinfo {volume} {779}\ (\bibinfo  {publisher} {Springer},\ \bibinfo {address} {Berlin, Heidelberg},\ \bibinfo {year} {2009})\BibitemShut {NoStop}%
\end{thebibliography}%

\end{document}